\documentclass[a4paper,twocolumn,11pt,accepted=2023-09-22]{quantumarticle}
\pdfoutput=1
\usepackage[utf8]{inputenc}
\usepackage[english]{babel}
\usepackage[T1]{fontenc}
\usepackage{amsmath}
\usepackage{hyperref}
\usepackage{tikz}
\usepackage{lipsum}
\usepackage{graphicx}
\usepackage{dsfont}
\usepackage{dcolumn}
\usepackage{bm}
\usepackage{lipsum}
\usepackage{color}
\usepackage{simplewick}
\usepackage{braket}
\usepackage{float}
\usepackage{multirow}
\usepackage{tikz}
\usepackage[utf8]{inputenc}
\usepackage{braket}
\usepackage[acronym,symbols,nogroupskip,nomain,nonumberlist,nopostdot,toc]{glossaries}
\hypersetup{
    colorlinks=true,
    citecolor=black,
    filecolor=black,
    linkcolor=black,
    urlcolor=black,
    linktocpage=true
}

\usepackage[normalem]{ulem}
\usepackage{makecell}
\usepackage{tikz-cd}
\usepackage{tkz-tab}
\usepackage{caption}
\usepackage{latexsym}
\usepackage{amssymb}
\usepackage{subcaption} 
\usepackage[qm]{qcircuit}
\usepackage{xr}
\hypersetup{
    colorlinks=true,
    citecolor=black,
    filecolor=black,
    linkcolor=black,
    urlcolor=black,
    linktocpage=true
}

\usepackage[acronym,symbols,nogroupskip,nomain,nonumberlist,nopostdot,toc]{glossaries}

\usetikzlibrary{automata,positioning}

\newcommand\rz{\textcolor{cyan}{\text{R}_\text{Z}}}
\newcommand\rx{\textcolor{cyan}{\text{R}_\text{X}}}
\newcommand\ry{\textcolor{cyan}{\text{R}_\text{Y}}}
\newcommand\z{\text{Z}}
\newcommand{\quotes}[1]{``#1''}
\newcommand{\bt}{\bm{\theta}}
\newcommand{\ansatz}{\textit{Ansatz}}
\newcommand{\ansatze}{\textit{Ans\"atze}}
\newcommand{\ie}{i.\,e.}

\newcommand{\eg}{e.\,g.}

\newcommand{\si}{supplementary information}

\newacronym{tdse}{TDSE}{time-dependent Schr\"odinger equation}
\newacronym{tdh}{TDH}{time-dependent Hartree}
\newacronym{mctdh}{MCTDH}{multi-configuration time-dependent Hartree}
\newacronym{vmcg}{vMCG}{variational multiconfigurational Gaussian}
\newacronym{ccs}{CCS}{coupled coherent states}
\newacronym{vte}{VTE}{variational time evolution}
\newacronym{tdvp}{TDVP}{time-dependent variational principle}
\newacronym{qft}{QFT}{quantum Fourier transform}
\newacronym{vqa}{VQA}{variational quantum algorithm}
\newacronym{pes}{PES}{potential energy surface}
\newacronym[firstplural=equations of motion (EOMs)]{eom}{EOM}{equation of motion}
\newacronym{dvr}{DVR}{discrete variable representation}
\newacronym{ld}{LD}{local diagonal}

\begin{document}

\title{Quantum algorithms for grid-based variational time evolution}

\author{Pauline J. Ollitrault}
\thanks{present address: QC Ware, Palo Alto, USA and Paris, France}
\affiliation{IBM Quantum, IBM Research -- Zurich, S\"aumerstrasse 4, 8803 R\"uschlikon, Switzerland}

\author{Sven Jandura}
\thanks{present address: ISIS (UMR 7006), University of Strasbourg, 67000 Strasbourg, France}
\affiliation{IBM Quantum, IBM Research -- Zurich, S\"aumerstrasse 4, 8803 R\"uschlikon, Switzerland}

\author{Alexander Miessen}
\affiliation{IBM Quantum, IBM Research -- Zurich, S\"aumerstrasse 4, 8803 R\"uschlikon, Switzerland}

\author{Irene Burghardt}
\affiliation{Institute of Physical and Theoretical Chemistry, Goethe University Frankfurt, Max-von-Laue-Str. 7, D-60438 Frankfurt/Main, Germany}

\author{Rocco Martinazzo}
\affiliation{Department of Chemistry, Università degli Studi di Milano, Via Golgi 19, 20133 Milan, Italy}
\affiliation{Istituto di Scienze e Tecnologie Chimiche “Giulio Natta”, CNR, Via Golgi 19, 20133 Milan, Italy}

\author{Francesco Tacchino}
\affiliation{IBM Quantum, IBM Research -- Zurich, S\"aumerstrasse 4, 8803 R\"uschlikon, Switzerland}

\author{Ivano Tavernelli}
\email{ita@zurich.ibm.com}
\affiliation{IBM Quantum, IBM Research -- Zurich, S\"aumerstrasse 4, 8803 R\"uschlikon, Switzerland}
\maketitle

\begin{abstract}
  The simulation of quantum dynamics calls for quantum algorithms working in first quantized grid encodings. 
Here, we propose a variational quantum algorithm for performing quantum dynamics in first quantization.
In addition to the usual reduction in circuit depth conferred by variational approaches, this algorithm also enjoys several advantages compared to previously proposed ones. 
For instance, variational approaches suffer from the need for a large number of measurements. 
However, the grid encoding of first quantized Hamiltonians only requires measuring in position and momentum bases, irrespective of the system size.
Their combination with variational approaches is therefore particularly attractive.
Moreover, heuristic variational forms can be employed to overcome the limitation of the hard decomposition of Trotterized first quantized Hamiltonians into quantum gates. 
We apply this quantum algorithm to the dynamics of several systems in one and two dimensions. 
Our simulations exhibit the previously observed numerical instabilities of variational time propagation approaches. 
We show how they can be significantly attenuated through subspace diagonalization at a cost of an additional $\mathcal{O}(MN^2)$ 2-qubit gates where $M$ is the number of dimensions and $N^M$ is the total number of grid points.
\end{abstract}

\section{Introduction}
Simulating quantum dynamics is of foremost importance to understand a multitude of chemical processes. 
Despite an impressive progress in the development of computational methods~\cite{lindh2020quantum,gatti2017applications,meyer2009multidimensional, curchod2018ab}, accurate calculations remain restricted to molecules with less than a few tens of collective degrees of freedom~\cite{gatti2014molecular,madsen2020mr,weir2020nonadiabatic,curchod2018ab}. 
An alternative path towards the efficient modeling of quantum dynamics is to switch towards a new computational paradigm.
In particular, quantum computers have the potential to simulate quantum systems in polynomial time and memory~\cite{feynman1999, tacchino2020advanced, miessen2023quantum}. 
Following Feynman's thesis, Wiesner~\cite{wiesner1996simulations} and Zalka~\cite{zalka1998simulating} first designed a framework to simulate molecular quantum dynamics on a digital quantum computer with a grid encoding of real space. 
This framework was then made concrete and applied to the simulation of several small nuclear quantum systems in rectangular or harmonic potentials~\cite{fan2012quantum, benenti2008, somma2015, ostrowski2015quantum, macridin2018electron}.
More recently, it was also extended to the simulation of the non-adiabatic dynamics of nuclear wavepackets~\cite{ollitrault2020non,Ollitrault2021molecular}.
All these approaches share the same circuit representation of the time evolution operator, obtained from Trotter approximation of the latter~\cite{berry2007}. 
However, in general, this leads to very deep quantum circuits, which greatly exceed the capacities of present quantum hardware due to their limited coherence times.
This is particularly true when the Hamiltonian is given in first quantization as, in this case, an efficient encoding of even a single Trotter step into a quantum circuit is cumbersome~\cite{ollitrault2020non, woerner2019quantum, haner2018optimizing}.
As a consequence, the type of implementable potentials is restricted. 
For instance, all the works cited above only present dynamics under potentials that can be defined with a polynomial function of the position.
It is worth emphasising here that active research is undergoing to implement Coulombic potentials~\cite{kassal2008, chan2022gridbased, jones2012, haner2018optimizing}.

To address the issue of circuit depth, \gls{vte} quantum algorithms were proposed. For a comprehensive overview see for instance Refs.~\cite{yuan2019, cerezo2021}. 
Relying on an iterative exchange of information between a classical and a quantum computer, these algorithms allow to work with shallower quantum circuits of constant depth in time. 
In particular, Li and Benjamin~\cite{li2017} first showed how to use a variational principle to simulate the real time dynamics of quantum systems and applied it to a quantum Ising model. 

The goal of the present work is to extend this approach to the simulation of wavepacket quantum dynamics in a grid-based encoding for general Hamiltonians given in first quantization.
The impact of the resulting algorithm lies beyond the ability of performing simulations on noisy near-term quantum hardware as it enables grid-based quantum dynamics in the first place, which can be a highly non-trivial task with a Trotter approximation.
We highlight the emergence of strong numerical instabilities in this context and present a local diagonalization scheme for making the variational time evolution algorithm stable and efficient.
We expect this novel approach to be easily extendable and beneficial to any problem instance similarly discretized, as is the case for example in non-linear problems~\cite{lubasch2020variational}, quantum field theory~\cite{macridin2022bosonic}, electron-phonon systems~\cite{macridin2018electron}, or quantum risk analysis~\cite{woerner2019quantum}.\\

\section{Wavepacket dynamics in position space}
In this work, we study the wavepacket dynamics governed by the \gls{tdse}
\begin{equation}
i\hbar \frac{\partial}{\partial t} \psi(x,t)=H(x) \, \psi(x,t) 
\label{td_SE}
\end{equation}
where $x$ is a spacial dimension mapped here to a one or two-dimensional grid. 
The wavefunction is normalized, $\int |\psi(x,t)|^2 \, dx =1$, so that the $|\psi(x,t)|^2$ becomes the probability probability for finding the system at position $x$ of the grid at time $t$. The time variable is considered as a continuous parameter. 
The Hamiltonian in Eq.~\eqref{td_SE} is given by
\begin{equation}
H(x)=-\frac{\hbar^2}{2m} \nabla^2 + V(x)
\label{Hamil_pos_rep}
\end{equation}
where the first term is the kinetic energy of the system and the second term describes a static external potential. 
In this framework, the time-evolution operator takes the form
\begin{equation}
U(t)=e^{-iHt/\hbar} \, .
\end{equation}
For an initial wavepacket $\psi(x,t_0)$ (mostly not an eigenstate of the Hamiltonian) the time evolution is given by
\begin{equation}
\psi(x,t)=e^{-iH(t-t_0)/\hbar} \,  \psi(x,t_0) 
\end{equation}
which can be evaluated numerically by means of standard integration schemes (such as the Trotter decomposition).
Note that the kinetic operator (first term in Eq.~\eqref{Hamil_pos_rep}) becomes diagonal in the momentum representation, i.e., after applying the Fourier transformation: $x\rightarrow{p=-i \hbar \partial/\partial x}$; it is therefore a common practice to define integrators of the time-evolution evolution operator, which leverage the advantages of both position and momentum representations of the system Hamiltonian. 

Alternatively, one can also express the time evolution through a variational principle.
We will delve deeper into this concept in the upcoming section.

\section{VTE in grid encodings}
The variational approach to quantum dynamics aims to approximate the solution of the \gls{tdse} on a low-dimensional submanifold of the full Hilbert space.
The trial wavefunction defined on this manifold $\ket{\psi(\bm{\theta}(t))}$ is parameterized by a set of $n_p$ time-dependent parameters, $\bt \equiv \bt(t) = \{ \theta_1(t), ..., \theta_{n_p}(t)\}$.
For a given Hamiltonian $\mathcal{H}$, the McLachlan variational principle~\cite{yuan2019, yao2021, gomes2021, mcardle2019variational, hackl2020} leads to the following equations of motion
\begin{equation}
    \text{F}\dot{\bt}=\text{V} \ ,
    \label{eq:vte}
\end{equation}
with 
\begin{equation*}
F_{kj} = \Re \bigl( \braket{\partial_{\theta_k }\psi | \partial_{\theta_j}\psi} -  \braket{\partial_{\theta_k}\psi | \psi} \braket{\psi|\partial_{\theta_j}\psi} \bigr) 
\end{equation*}
and
\begin{equation*}
V_k = \Im \bigl( \bra{\partial_{\theta_k}\psi}\mathcal{H}\ket{\psi} - \braket{\partial_{\theta_k}\psi|\psi}\bra{\psi}\mathcal{H}\ket{\psi} \bigr) \, .
\end{equation*}
Further details on these equations are available in the \si. 

The calculation of the $F_{kj}$ and $V_k$ matrix and vector elements are classically hampered by the high dimension of $\ket{\psi(\bt)}$ which ensures accurate simulations.
Instead, they can be efficiently measured on a trial wavefunction encoded in the state of a qubit register. 
In this case, the trial wavefunction is defined as \begin{equation} 
\ket{\psi(\bt)} = \mathcal{U}(\bt)\ket{\phi}
\end{equation} 
where $\ket{\phi}$ is a reference state and $\mathcal{U}(\bt)$ is a unitary operator (e.g., the quantum circuit) depending on real parameters $\bt \equiv \bt(t)$.
The measurement of analytic gradients on quantum computers was discussed in Ref.~\cite{Schuld2019gradients}. 
If the parameters $\bt$ are chosen as the rotation angles of single qubit gates, there always exists a simple circuit $\mathcal{W}(\bt)$ such that \begin{equation} 
\ket{\partial_{\theta_k}\psi(\bt)} = -\frac{i}{2} \mathcal{W}(\bt)\ket{\phi} \, .
\end{equation} 
In the \si{} we recall the canonical approach to computing $F_{kj}$ and $V_k$ when the Hamiltonian is written as a weighted sum of Pauli tensor strings, $\mathcal{H}=\sum_p c_p \mathcal{P}_p$. 

Here, instead, we will focus on Hamiltonians expressed in first quantization, $\mathcal{H} = \bm{p}^2 / 2m + \mathcal{V}(\bm{r})$,
where $\bm{p}$ is the momentum, $m$ the mass, and $\mathcal{V}(\bm{r})$ the potential given as a function of the position $\bm{r}$. 
$\mathcal{H}$ describes an $M$-dimensional systems.
The time-evolution is directly performed in momentum and position space, discretized on a grid. 
The $N$ points, per dimension, of the grid are encoded in the basis states of $N_q=\log_2(N)$ qubits. 
The total number of qubits is then $MN_q$.
In this case, no explicit transformation to a basis representation is required, saving the numerical effort associated with such computations, which would scale as the square of the basis set size~\cite{manthe2002quantum}.
Note that the encoding (real space discretization on a grid) employed here is different from the \gls{dvr} of Ref.~\cite{lee2021variational}.
We argue in favor of the present grid encoding as the \gls{dvr} representation leads to an exponentially growing number of Hamiltonian terms. 
In our case, the expectation value of the Hamiltonian is simply obtained from two sets of measurements, one in the momentum basis and one in the position basis.
In fact, this constitutes an important advantage compared to previous implementations of the same \gls{vte} quantum algorithm, \eg, in second quantization or in the \gls{dvr} representation. Indeed, the high number of measurements required to perform the \gls{vte} becomes quickly impractical when increasing the dimensionality of the problem~\cite{miessen2021, gacon2021simultaneous, barison2021efficient}.

This advantage appears in the calculation of the right-hand side of Eq.~\ref{eq:vte}, which requires measuring 
$
\bra{\psi(\bt)}\mathcal{H}\ket{\psi(\bt)}
$ 
and 
$
\Im(\bra{\partial_{\theta_k}\psi(\bt)}\mathcal{H}\ket{\psi(\bt)})
$.
The first term is straightforward and reduces to measuring the expectation values $\bra{\psi(\bt)}\bm{p}^2\ket{\psi(\bt)}$ and $\bra{\psi(\bt)}\mathcal{V}(\bm{r})\ket{\psi(\bt)}$. 
In practice, this is obtained by repeatedly preparing and measuring the state $\ket{\psi(\bt)}$ in the computational (position) basis. 
For each outcome of the binary representation of $\bm{r}$, $\mathcal{V}(\bm{r})$ is classically computed and $\bra{\psi(\bt)}\mathcal{V}(\bm{r})\ket{\psi(\bt)}$ is obtained by averaging over all realizations of $\mathcal{V}(\bm{r})$. 
The same is done for the kinetic term by measuring in the momentum basis, \ie, applying a \gls{qft} right before the measurement (see Ref.~\cite{ollitrault2020non} and \si). 

The second term in $V_k$ can be rewritten as
\begin{align*}
    \Im\big(\bra{\partial_{\theta_k}\psi(\bt)}\mathcal{H} & \ket{\psi(\bt)}\big) = \notag \\ 
    &1/2\, \Re\big(\bra{\phi}\mathcal{W}(\bt)^{\dagger}\mathcal{H}\mathcal{U}(\bt)\ket{\phi}\big) \, .
\end{align*}
The scalar $\Re\big(\bra{\phi}\mathcal{W}^{\dagger}(\bt)\mathcal{V}(\bm{r})\mathcal{U}(\bt)\ket{\phi}\big)$ is then obtained from the calculation of $\mathbb{E}[(-1)^s\mathcal{V}(r)]$ with $r$ and $s$ the measurement outcomes of the quantum register and ancilla qubit, respectively, obtained from the following quantum circuit
\begin{equation*}
    \Qcircuit @C=0.5em @R=0.5em @!R {
	 	\lstick{  \ket{0} } &\gate{H} & \ctrl{1} & \gate{X} & \ctrl{1} & \gate{X} & \gate{H} & \meter\\
	 	\lstick{ \ket{\phi} } & \qw {/} & \gate{\mathcal{W}(\bt)} & \qw & \gate{\mathcal{U}(\bt)} & \qw & \qw  & \meter  }.
\end{equation*}
The proof is given in the \si. 
The same is done for the kinetic term by measuring in the momentum basis, \ie, applying a \gls{qft} right before the measurement. \\

\section{The wavefunction representation and quantum advantage}
%
The quantum advantage of the approach presented herein relies on the possibility to accurately approximate the wavefunction with a quantum computer, using a number of variational parameters, $n_p$, that is much smaller than the size of the full Hilbert space.
The parameters can then be propagated efficiently according to Eq.~\eqref{eq:vte}, as long as $n_p$ scales favorably (i.e., polynomially) with the system size.

When working in first quantization, the translation of a Trotter operator $\mathcal{T}$ into a quantum circuit can be very costly in terms of the gate count.
Indeed, an efficient decomposition of 
\begin{equation} 
\mathcal{T} = \exp(-i\mathcal{V}(\bm{r})t)
\end{equation}
for an arbitrary potential energy function $\mathcal{V}(\bm{r})$ into a set of quantum gates is not guaranteed to exist. In fact, the quantum circuit for encoding a general function requires either exponentially many gates or ancilla qubits to perform quantum arithmetic. In the latter case, the targeted function is generally approximated and the number of ancilla qubits scales exponentially with the inverse of the desired accuracy~\cite{mitarai2019, woerner2019quantum, haner2018optimizing, ollitrault2020non}.

On one hand, this speaks in favor of the use of variational approaches for time-propagation over Trotter-based algorithms since the former does not require the implementation of an accurate (and costly) gate decomposition of the operator $\mathcal{T}$. 
On the other hand, finding appropriate, physically motivated, variational \ansatze{} is still a challenge, and heuristic, hardware efficient approaches are therefore preferred. 
Hence, as a first demonstration in low dimensions, we will limit ourselves to heuristic variational forms, which can be implemented on quantum hardware. 
These can be systematically improved by repeating the same quantum circuit units with independent parameters, \ie, by increasing the circuit depth $d$. 
The different heuristic circuits employed in this work are detailed in the \si. \\
\begin{figure}[t]
    \centering
    \includegraphics[width = 0.9 \columnwidth]{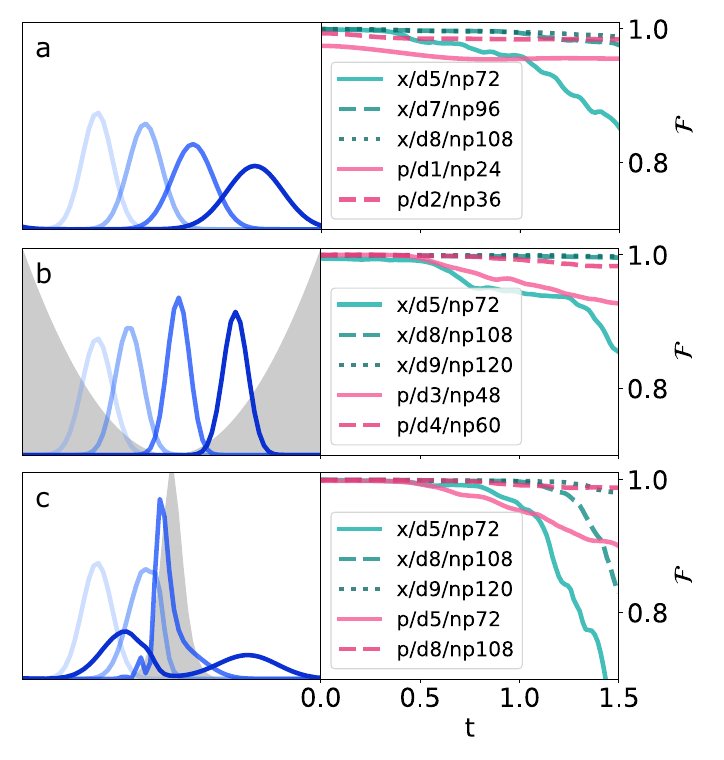}
    \caption{\gls{vte} dynamics for the three one-dimensional systems considered in this study.
    \textit{left panels} Snapshots of the modulus square of the exact wavefunctions at times $t=0.00, 0.45, 0.91$, and $1.5$ (lightest to darkest curve, respectively) for a) the free particle, b) the harmonic oscillator, c) the Eckart barrier. 
    \textit{right panels} Fidelity, $\mathcal{F}$ as a function of time, $t$, of the \gls{vte} dynamics in position (x) and momentum (p) space for a) the free particle, b) the harmonic oscillator, c) the Eckart barrier. The results are obtained with 6 qubits and variational form vf1 at depth $d$ (with corresponding number $n_p$ of variational parameters).}
    \label{fig:main_fig_1d}
\end{figure}

\section{Simulations in position and momentum spaces}
In the following, we study the performance of the quantum algorithm for concrete applications in first quantization. 
We seek to perform this study without introducing any quantum hardware noise bias.
Hence, the simulations are obtained in a perfect classical emulator of a quantum computer.
The expectation values are computed from matrix-vector multiplications. 
For this reason, we approximate the wavefunction derivatives with forward finite differences of step-size $10^{-8}$ (which should not be confused with the adaptive time-step used for the time propagation). We show in the \si{} that this step-size leads to converged results.
The equations of motion (Eq.~\eqref{eq:vte}) are solved with a least-squares approach as implemented in \textsc{NumPy}~\cite{numpy} with a cutoff ratio for small singular values, or reconditioning number, arbitrarily set to $10^{-6}$.
To solve the ordinary differential equations, we employ a state-of-the-art adaptive Runge-Kutta solver of order 5(4) available in \textsc{SciPy}-routines \cite{2020SciPy-NMeth}.
We first test the \gls{vte} approach in one-dimensional systems defined on a grid as described in Ref.~\cite{ollitrault2020non} and in the \si. 
The length of the box is $\text{L} = 14$. 
We study three different problems: a freely moving particle, a particle in a harmonic potential (harmonic oscillator), and a particle colliding with an Eckart barrier. 
The Hamiltonians for these three systems are given by 
\begin{align}
   \mathcal{H}_{\text{FP}} &= p^2/2m \, ,  \\\mathcal{H}_{\text{HO}} &= p^2/2m + c_1 x^2 \, , \\
   \mathcal{H}_{\text{EB}} &= p^2/2m + c_2/\cosh^2(c_3 x) \, .
\end{align}
We work in atomic units, taking $m=1$, $c_1 = 1$, $c_2 = 13$, and $c_3 = 3/2$.
In all cases, the state is initialized to a Gaussian wavepacket, 
\begin{equation}
\psi_0(x) = \mathcal{A} \, e^{\big(-1/4\big[(x-x_0)/\mathcal{B}\big]^2\big)} e^{(ip_0x)} \, ,
\end{equation}
with $\mathcal{A}$ the normalization factor, $\mathcal{B}$ the width of the wavepacket, and $p_0$ and $x_0$ the initial momentum and position, respectively.
The evolution is carried out for a total time $t_{\text{tot}} = 1.5$. 
Snapshots of the exact time evolution (from exact exponentiation of the Hamiltonian matrix) for each of the three systems are shown in Fig.~\ref{fig:main_fig_1d} (left panels) for times $t=0, 0.45, 0.91$, and $1.5$.

The space is discretized with $N_q=6$ qubits (corresponding to 64 grid points).
In this case, the full Hilbert space can be described by $n_p^{\text{full}}=2(2^6 - 1) = 126$ real parameters. 
It is clear that a quantum advantage can only be achieved with a much smaller number of variational parameters, satisfying $n_p \ll n_p^{\text{full}}$. 
We perform \gls{vte} simulations for each of the three aforementioned systems with the variational form vf1 (see \si) for different depths (or circuit repetitions, see \si).

The initial parameters $\bt_0$ are found by maximizing the fidelity 
\begin{equation}
\mathcal{F}(t=0) = |\braket{\psi_0 | \mathcal{U}(\bt_0) | 0}|^2 \, ,    
\end{equation}
where $\ket{0}$ is the vacuum state. 
Note that this procedure, which is sub-optimal in general, is implemented here for convenience. An extensive discussion on state initialization goes beyond the scope of this work and is left for future investigations.
The initial conditions are $(x_0, p_0) = (-3.5, 5)$ for the free particle and the Eckart barrier, and $(x_0, p_0) = (-3.5, 2)$ for the harmonic oscillator.
In all cases, the width of the initial wavepacket is set to $\mathcal{B}=1/\sqrt{2}$.

The fidelities 
\begin{equation}
\mathcal{F}(t) = |\braket{\psi_0 | e^{i\mathcal{H}t} \mathcal{U}(\bt(t))| 0}|^2 \, , 
\end{equation}
as a function of the simulation time, are shown in Fig.~\ref{fig:main_fig_1d} (right panels). 
These results demonstrate that, in general, the number of variational parameters required to maintain a fidelity above 95\% throughout the entire simulation time always approaches $n_p^{\text{full}}$, particularly so when tackling hard problems such as the scattering off an Eckart barrier, reducing the possibility for quantum advantage.
The evolution of all parameters is also given in the \si{} and shows sharp changes in their trajectories for each of the three systems. 
To our understanding, supported by the detailed study given in the \si, the above observations can be rationalized as follows. 
The chosen heuristic variational forms have enough flexibility to accurately and efficiently, \ie, with few variational parameters, approximate the targeted wavefunctions at all times of their dynamics.
Therefore, the loss of accuracy observed throughout the different simulations is not due to the variational forms, but is an inherent effect of the \gls{vte}.
When the number of variational parameters is insufficient, the dynamics strongly depends on the numerical setup of the simulation (grid mesh, reconditioning number, initial parameters, etc.). 
These observations are in agreement with Ref.~\cite{lee2021variational} where strong numerical instabilities were also put forward.
Surprisingly, the correct dynamics are always recovered when increasing the number of variational parameters.
In this case, the algorithm is stable.
In addition, we observe that the more complex the dynamics gets, the more parameters are needed to obtained a robust and stable dynamics. Here, the term \textit{complex} refers to systems with a large support in position and frequency spaces in their canonical representation, which lack of any particular symmetry reduction property.
For completion, we also show in the \si{} the robustness of the variational approach to the introduction of hardware noise.

We repeated our simulations in the momentum representation by defining a variational \ansatz{} of the form 
\begin{equation}
    \mathcal{U}_p(\bt) = \text{QFT} \, \mathcal{U}(\bt) \, . 
\end{equation}
The results are given in Fig.~\ref{fig:main_fig_1d}.
As expected, the evolution of the free particle can now be performed accurately with very few parameters, since the introduced momentum basis diagonalizes the Hamiltonian.
Interestingly however, the evolution is also smoother in the case of the harmonic oscillator and the Eckart barrier.
Note that we do not expect the harmonic oscillator to be symmetric in position and momentum representations, since the position and momentum grids are different and the corresponding potentials come with different prefactors. 
In the \si{}, we show that this momentum space representation improves the dynamics compared to the position space in all cases considered in this work. 
Results obtained by combining both the position and the momentum representation in a common \ansatz{} are also given in the \si{} but do not show substantial improvements. 

In order to search for efficient heuristic \ansatze{}, we systematically change the characteristics of the variational circuit, namely the single-qubit rotations, the type of entangling gates, and the connectivity of the entangling block (see \si{} for details).
We apply all these different variational forms to the simulation of the wavepacket interacting with the Eckart barrier.
The results presented in the \si{} show that, in general, the accuracy is improved by working in the momentum representation and increasing the circuit depth.
However, no clear trend for the design of more efficient variational forms could be identified.\\

\section{Simulations in local diagonal space}
The introduction of unitary transformations that achieve a (at least partial) diagonalization of the Hamiltonian seems essential to make the quantum \gls{vte} efficient in the grid representation. The reasoning behind this approach is that a significant part of the dynamics can be associated with the eigenstate evolution in uncoupled subspaces, such that only the remaining, often smooth evolution needs to be treated in a variational setting. We elaborate on this concept in the \si.

In the \si, we show some preliminary one-dimensional tests demonstrating that an approximate diagonalization of the system Hamiltonian can already give a sufficiently accurate and stable time evolution.
Of more interest is the application of this approach to multi-dimensional systems. 
Following our previous observations, we expect that,
for a multi-dimensional configuration space, an improvement can already be achieved by diagonalizing each dimension independently, \ie, by fixing all but one coordinate of the Hamiltonian. This leads to a quantum circuit of the form
\begin{equation*}
    \Qcircuit @C=0.5em @R=0.5em @!R {
	 	\lstick{\ket{0}^{\otimes N_q}} & \qw {/} &\multigate{3}{\mathcal{U}(\bm{\theta})}& \gate{\mathcal{D}_1} & \qw \\
        \lstick{\ket{0}^{\otimes N_q}} & \qw {/} & \ghost{\mathcal{U}(\bm{\theta})}& \gate{\mathcal{D}_2} & \qw\\
            & & & \vdots &\\
        \lstick{\ket{0}^{\otimes N_q}} & \qw {/} & \ghost{\mathcal{U}(\bm{\theta})}& \gate{\mathcal{D}_M} & \qw 
        }
	 	\label{eq:multidim_diag_circuit}
\end{equation*}
We call the space obtained from this basis transformation the \gls{ld} space. In the next paragraph, we discuss the scaling of this approach.

As stated earlier, we employ $N$ grid points, corresponding to $N_q = \log_2(N)$ qubits, for each of the $M$ dimensions.
Hence, the total number of qubits is $MN_q$.
The total size of the problem is $N^M$, prohibitively large for direct classical simulations in high dimensions. 
Efficient algorithms should scale to low polynomial order in $N$ and $M$. 
Our approach first requires the diagonalization of $M$ matrices of size $(N,N)$.
This step generally scales as $\mathcal{O}(N^3)$.
The resulting unitary is then translated into a quantum circuit.
The classical scaling of this operation is $\mathcal{O}(N_qN^3)$~\cite{Iten2019}, leading to $\mathcal{O}(N^2)$ CNOT gates~\cite{Iten2016}.
Finally, the \gls{vte} is performed with a variational form comprising $n_p$ parameters. The partial diagonalization step ensures the scaling of $n_p$ to be of low polynomial order, \ie, $\mathcal{O}\big((MN_q)^P\big)$ where $P$ is small enough such that $(MN_q)^P \ll N^M$ (see the \si{} for a more comprehensive description of the \gls{ld} basis and its effect on $n_p$). 
At each time step of the \gls{vte}, the equations of motion are reconstructed with $\mathcal{O}(n_p^2)$ measurements and solved classically at cost $\mathcal{O}\big(n_p^3\big)$.
Recent approaches such as the one of Ref.~\cite{gacon2021simultaneous} focus on reducing the scaling in $n_p$ for obtaining the M matrix and could be extended to the present case. 
Finally, we obtain an efficient hybrid quantum-classical time evolution algorithm with polynomial scaling in $N$ and $M$. 
\begin{figure}[t]
    \centering
    \includegraphics[width = \columnwidth]{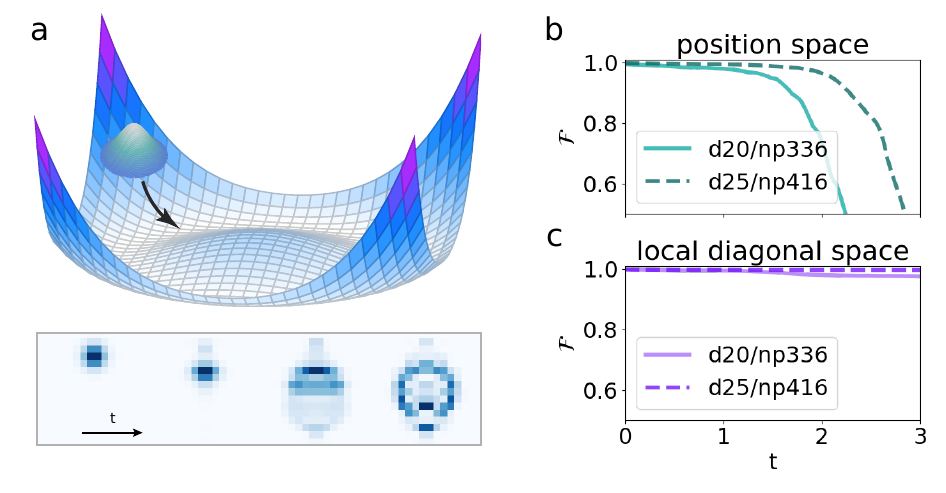}
    \caption{\gls{vte} dynamics of a two-dimensional system, representing a wavepacket evolving on a \quotes{mexican hat} potential.\\
    a) 3-dimensional representation of the system at the top and, at the bottom, snapshots of the modulus square of the exact wavefunction (darker color indicates larger amplitude) at times $t=0.00, 0.91, 1.8$, and $3.0$. \\
    Beside are the results of the \gls{vte} dynamics obtained in b) the position space and c) the LD space. The fidelity, $\mathcal{F}$ as a function of time, $t$, is obtained with 8 qubits and variational form vf1 at depth $d$ (with the corresponding number $n_p$ of variational parameters).}
    \label{fig:mh_results}
\end{figure}

As a proof of concept, we test this approach on the evolution of a two-dimensional system on a \textit{Mexican Hat} \gls{pes}. 
The Hamiltonian reads
\begin{equation}
    \mathcal{H}_{\text{MH}} = (p_x^2 + p_y^2)/2m + c_4 r^4 - c_5r^2 \, ,
\end{equation}
with $r^2 = x^2+y^2$, $c_4=0.1$, $c_5=1$, and again $m=1$.
The space is discretized with 8 qubits (4 per dimension), corresponding to a total of 256 grid points. 
Note that the Hilbert space can be fully represented by 510 real parameters.
The wavepacket is evolved in the region of space $x\in[-5,5]$ and $y\in[-5,5]$.
It is initialized at $(x_0,y_0)=(-3.0, 0)$ with no initial momentum and width $\mathcal{B}_x=\mathcal{B}_y=1/\sqrt{2}$.  
The momentum increases in subsequent time steps as the wavepacket slides down the brim of the \gls{pes}. 
A graphical representation of the system is given in Fig.~\ref{fig:mh_results}a together with snapshots of the exact evolution for times $t=0, 0.91, 1.8$, and $3.0$.

We simulate the quantum dynamics with \gls{vte} up to $t_{\text{tot}}=3.0$, employing the same settings as before, and using both the position and the \gls{ld} \ansatz.
The one-dimensional Hamiltonians $\mathcal{H}_x = \mathcal{H}(x, 0)$ and $\mathcal{H}_y = \mathcal{H}(0, y)$ are diagonalized to get the quantum circuits $\mathcal{D}_x$ and $\mathcal{D}_y$. 
For both the evolutions in position and \gls{ld} space, the parameterized part of the quantum circuit $\mathcal{U}(\bm{\theta})$ corresponds to the variational form vf1 with depth 20 and depth 25 (336 and 416 parameters, respectively).
The results are displayed in Fig.~\ref{fig:mh_results}b and c.
Significant improvements are obtained in \gls{ld} space.
Note that these results were obtained from a choice of one-dimensional Hamiltonians that are very simple to diagonalize.
They could be improved even further by exploiting symmetries of the system or by diagonalizing low-dimensional mean-field Hamiltonians.\\

\section{Conclusions}
We introduced a quantum \gls{vte} algorithm to perform nuclear wavepacket dynamics on a grid in first quantization. 
This variational algorithm manifests a crucial advantage compared to Trotter-like quantum approaches to this problem class, namely the fact that it does not require the direct implementation of the time-evolution operator (exponentiating the Hamiltonian) in the qubit register. 
Furthermore, we stressed the advantage of our method in relation to the need of sampling expectation values in only two bases representations (position and momentum), irrespective of the system size.
We studied the performance of the quantum algorithm in classical emulations for several one- and two-dimensional systems.
In general, we observed strong numerical instabilities when performing the dynamics with an efficient number of variational parameters.
However, we could demonstrate that the accuracy of the quantum algorithm can be improved by expressing the variational quantum circuit in a problem specific basis.
This basis is obtained by diagonalizing each dimension of the system independently, without introducing significant computational overhead.
Moreover, adaptive approaches for variational time evolution~\cite{yao2021} might help to further reduce ansatz cost in situations where our diagonalization procedure is not applicable.
However, selecting a suitable operator pool for constructing the adaptive ansatz could be challenging, since these approaches typically rely on the Trotterization of the time evolution operator.
As mentioned previously, this is not generally efficient in first quantization.

We discussed the overall cost of the proposed approach, which shows an effective polynomial scaling in both the number of grid points per dimension, $N$, and the number of system dimensions, $M$ (for a total of $N^M$ grid points).
It is worth noting that, treating higher dimensional grids, the qubit connectivity of the respective quantum hardware will have significant impact on implementation details of our approach and the ordering of qubits will have to be taken into consideration.
Going further, this approach can be easily extended to the simultaneous treatment of electronic and nuclear degrees of freedom, opening up new opportunities for the simulation of non-adiabatic dynamics beyond the Born-Oppenheimer approximation. When dealing with identical particles, the system wavefunction needs to be adapted accordingly:  antisymmetrized for fermions and symmetrized for bosons upon particle exchange. 
The generation of quantum circuits fulfilling invariance under (anti-)symmetrization is discussed in detail in the literature~\cite{chan2022gridbased}.\\

\section{Acknowledgements}
The authors thank Christa Zoufal for useful discussions and acknowledge financial support from the Swiss National Science Foundation (SNF) through the grant No. 200021-179312.\\
IBM, the IBM logo, and ibm.com are trademarks of International Business Machines Corp., registered in many jurisdictions worldwide. Other product and service names might be trademarks of IBM or other companies. The current list of IBM trademarks is available at https://www.ibm.com/legal/copytrade.

\bibliographystyle{quantum}
\bibliography{2022_VTE_1Q/varqd.bib}

\clearpage
\onecolumn
\appendix

\setcounter{equation}{0}
\setcounter{section}{0}
\setcounter{figure}{0}
\setcounter{table}{0}
\setcounter{page}{1}
\renewcommand{\theequation}{S\arabic{equation}}
\renewcommand{\thefigure}{S\arabic{figure}}
%
%
\section{The grid encoding}

In this work, real space is discretized on a grid which points are mapped to the basis states, $\ket{j}$, of the quantum register. 
We illustrate the grid encoding of this work in a one dimensional example. 
The length of the grid is $L$. There are $N$ points and $N_q = \log_2(N)$ qubits. 
The initial point in position space is $x_0 = -L/2$. Each subsequent point is defined as $x_j = x_0 + \frac{L}{N-1} j$. $j$ is an integer which binary representation is given by $\ket{j}$.
In the reciprocal (momentum) space the points are then defined as $p_j = p_0 + \frac{2\pi}{L} j$ with $p_0 = - \frac{N\pi}{L}$.
This shift allows to account for nagtive values of the momentum.
This choice implies the use of a centered Quantum Fourier Transform (cQFT) operator to implement the switch from the position to the momentum space.
This cQFT can simply be implemented by adding a X gate on the last qubit right before and after the QFT (for cQFT, \ie{} $\text{cQFT}\equiv (I\otimes I \otimes ... \otimes X) \text{QFT} (I\otimes I \otimes ... \otimes X)$) and QFT$^{-1}$ (for cQFT$^{-1}$) such that they respectively undergo a cyclic permutation. The equation defining this permutation is given below for QFT as an illustration:
\begin{equation}
     \text{\textbf{cQFT}} =
    \begin{pmatrix}
     & & &1 & \cdots & 0\\
     &\cdots& & & \ddots & \\ 
     & & & 0 & \cdots & 1 \\
     1 & \cdots & 0 & & &\\
     & \ddots & & &\cdots&\\ 
     0 & \cdots & 1 &&& \\
    \end{pmatrix}
    \text{\textbf{QFT}}.
\end{equation}

\section{The time-dependent variational principle in first and second quantization}

The variational approach to quantum dynamics aims to approximate the solution of the \gls{tdse} on a low-dimensional submanifold of the full Hilbert space.
The trial wavefunction defined on this manifold $\ket{\psi(\bm{\theta}(t))}$ is parameterized by a set of $n_p$ time-dependent parameters, $\bt(t) = \{ \theta_1(t), ..., \theta_{n_p}(t)\}$.
A \gls{tdvp} defines the optimal evolution of the parameters within the submanifold of the full Hilbert space. 
There exist different formulations of the \gls{tdvp}.
For K\"ahler manifolds, \ie, when the tangent space is a complex subspace~\cite{hackl2020}, all the different formulations lead to the same equations of motion for the variational parameters. 
However, this is not the case for unitary parameterizations of the type
\begin{equation}
    \ket{\psi(\bt(t))} = \mathcal{U}(\bt(t))\ket{\phi},
    \label{eq:trial_wf}
\end{equation}
where $\ket{\phi}$ is a reference state and $\mathcal{U}(\bm{\theta}(t))$ is a unitary operator (e.g., the quantum circuit) depending on real parameters $\bm{\theta}(t)$~\cite{hackl2020}. In this case, the equations of motion differ and hold distinct properties such as the conservation of the norm or the energy.
Recent works~\cite{yuan2019, yao2021, gomes2021, mcardle2019variational, hackl2020} promoted the use of the equations of motion derived from the McLachlan variational principle due to of their higher numerical stability~\cite{hackl2020}.
For a given Hamiltonian $\mathcal{H}$, these equations, when accounting for a global phase mismatch (see Ref.~\cite{yuan2019} for a thorough derivation), read  
\begin{equation}
    \text{F}\dot{\bt}=\text{V} \ ,
    \label{eq:vte}
\end{equation}
with 
\begin{equation}
    F_{kj} := \Re \Bigl( \braket{\partial_{\theta_k }\psi | \partial_{\theta_j}\psi} -  \braket{\partial_{\theta_k}\psi | \psi} \braket{\psi|\partial_{\theta_j}\psi} \Bigr) \ ,
    \label{eq:fubini_study}
\end{equation}
and 
\begin{equation}
    V_k := \Im \Bigl( \bra{\partial_{\theta_k}\psi}\mathcal{H}\ket{\psi} - \braket{\partial_{\theta_k}\psi|\psi}\bra{\psi}\mathcal{H}\ket{\psi} \Bigr) \ .
    \label{eq:projection}
\end{equation}
For a regular parametrization, Eq.~\eqref{eq:fubini_study} represents a quantum metric (the Fubini-Study metric) in parameter space. In conjuction with Eq.~\eqref{eq:fubini_study}, Eq.~\eqref{eq:projection} shows that Eq.~\eqref{eq:vte} results from the orthogonal projection of the exact time-derivative of $\ket{\psi}$ on the variational manifold. Note that the variational manifold is entirely general, and any parametrized form of the wave function can be used to define it, provided that the parametrization is regular.\\

The calculation of the $F_{kj}$ and $V_k$ matrix and vector elements are classically hampered by the high dimension of $\ket{\psi(\bt)}$.
Instead, they can be efficiently measured on a trial wavefunction encoded in the state of a qubit register. 
In this case, the trial wavefunction $\ket{\psi(\bt)}$ (see Eq.~\eqref{eq:trial_wf}) is defined by the variational parameters $\bt$, which enter the quantum circuit $\mathcal{U}$ as rotation angles of the type $R_X(\theta_k) = e^{-i\theta_kX/2}$, $R_Y(\theta_k) = e^{-i\theta_kY/2}$, or $R_Z(\theta_k) = e^{-i\theta_kZ/2}$. Here, $X$, $Y$, and $Z$ are the Pauli matrices. 
The measurement of analytic gradients on quantum computers was discussed in Ref.~\cite{Schuld2019gradients}. 
Here, we detail the technical differences between the algorithm for Hamiltonians defined as a sum of Pauli operators and for first quantized Hamiltonians encoded on a grid.

Consider the following general form for a variational quantum circuit $\mathcal{U}(\bt(t))\equiv \mathcal{U}(\bt)$ comprising $n_p$ parameters and acting on $N_q$ qubits
\begin{equation*}
    \qquad \qquad \Qcircuit @C=0.4em @R=0.5em @!R {
	 	\lstick{  \ket{0} } & \qw & \multigate{1}{\mathcal{U}_1(\theta_1 ... \theta_{k-1})} & \gate{R_{\sigma}(\theta_k)} & \multigate{1}{\mathcal{U}_2(\theta_{k+1} ... \theta_{n_p})} \\
	 	\lstick{ \ket{0}^{\otimes N_q-1} } & \qw {/} & \ghost{\mathcal{U}_1(\theta_1 ... \theta_{k-1})} & \qw & \ghost{\mathcal{U}_2(\theta_{k+1} ... \theta_{n_p})}\\
	 	}
\end{equation*}
with $\sigma \in \{X, Y, Z\}$.
Then, $\partial_{\theta_k} \mathcal{U}(\bt)$ can be computed from the following quantum circuit 
\begin{equation*}
    \qquad \qquad \Qcircuit @C=0.4em @R=0.5em @!R {
	 	\lstick{  \ket{0} } & \qw & \multigate{1}{\mathcal{U}_1(\theta_1 ... \theta_{k-1})} & \gate{\sigma} & \gate{R_{\sigma}(\theta_k)} & \multigate{1}{\mathcal{U}_2(\theta_{k+1} ... \theta_{n_p})} \\
	 	\lstick{ \ket{0}^{\otimes N_q-1} } & \qw {/} & \ghost{\mathcal{U}_1(\theta_1 ... \theta_{k-1})} & \qw & \qw & \ghost{\mathcal{U}_2(\theta_{k+1} ... \theta_{n_p})}
	 	}
\end{equation*}
All matrix elements of F and V can therefore be written as $\bra{\phi} \mathcal{W}_1^{\dagger} \mathcal{O} \mathcal{W}_2 \ket{\phi}$, where $\mathcal{W}_1 \equiv \mathcal{W}_1(\bt)$ and $\mathcal{W}_2 \equiv \mathcal{W}_2(\bt)$ are unitary quantum circuits, and $\mathcal{O}$ is a general operator.

We start by considering the system Hamiltonian in second quantization. 
After mapping the second quantized operators to Pauli operators~\cite{mcardle2020review, miessen2021}, the Hamiltonian is written as a weighted sum of Pauli tensor strings, $\mathcal{H}=\sum_p c_p \mathcal{P}_p$. 
With $\mathcal{O} = \mathcal{H}$, the general form of the expectation value given above becomes
\begin{equation}
    \bra{\phi} \mathcal{W}_1^{\dagger} \sum_p c_p \mathcal{P}_p \mathcal{W}_2 \ket{\phi} = \sum_p c_p \bra{\phi} \mathcal{W}_1^{\dagger} \mathcal{W}_{2,p} \ket{\phi} \ ,
\end{equation}
where $\mathcal{W}_{2,p} = \mathcal{P}_p \mathcal{W}_2$.
Each term can then be efficiently measured as~\cite{somma2002}
\begin{equation}
\begin{split}
    \bra{\phi} \mathcal{W}_1^{\dagger} &\mathcal{W}_{2,p} \ket{\phi} \equiv \\\qquad
    &\Qcircuit @C=0.5em @R=0.5em @!R {
	 	\lstick{  \ket{0} } & \gate{H} & \ctrl{1} & \gate{X} & \ctrl{1} & \gate{X} & \qw & \qw & \rstick{\!\!\!\braket{2\sigma_+}} \\
	 	\lstick{ \ket{\phi} } & \qw {/} & \gate{\mathcal{W}_{2,p}} & \qw & \gate{\mathcal{W}_1} & \qw & \qw & \qw & \rstick{}\\
	 	}\qquad
\end{split}
	 	\label{eq:laflamme_circ}
\end{equation}
The circuit steps can be written as follows (we write $\ket{\phi} \ket{\varphi}$ for $\ket{\phi} \otimes \ket{\varphi}$, single-qubit kets $\ket{0}$ denote the ancilla qubit, while multi-qubit states $\ket{\phi}$ denote the qubit register encoding the grid):
\begin{enumerate}
    \item $\ket{\phi}\ket{0} \longrightarrow \frac{1}{\sqrt{2}}\ket{\phi}\ket{0} + \frac{1}{\sqrt{2}}\ket{\phi}\ket{1}$
    \item $\frac{1}{\sqrt{2}}\ket{\phi}\ket{0} + \frac{1}{\sqrt{2}}\ket{\phi}\ket{1} \longrightarrow \frac{1}{\sqrt{2}}\ket{\phi}\ket{0} + \frac{1}{\sqrt{2}}\mathcal{W}_{2,p}\ket{\phi}\ket{1}$ 
    \item $\frac{1}{\sqrt{2}}\ket{\phi}\ket{0} + \frac{1}{\sqrt{2}}\mathcal{W}_{2,p}\ket{\phi}\ket{1} \longrightarrow \frac{1}{\sqrt{2}}\mathcal{W}_1\ket{\phi}\ket{0} + \frac{1}{\sqrt{2}}\mathcal{W}_{2,p}\ket{\phi}\ket{1}$ 
    \item Measuring the expectation value of $\sigma_+ = \ket{0}\bra{1}$ on the ancilla qubit leads to
\end{enumerate}
\begin{equation}
        \braket{\sigma_+} = \frac{1}{2} \bigg( \bra{1}\bra{\phi} \mathcal{W}_{2,p}^{\dagger} + \bra{0}\bra{\phi} \mathcal{W}_1^{\dagger}\bigg) \ket{0}\bra{1} \bigg( \mathcal{U}\ket{\phi}\ket{0} + \mathcal{W}_{2,p}\ket{\phi}\ket{1} \bigg) \ .
\end{equation}
Since $\braket{1|0} = \braket{0|1} = 0$, we obtain
\begin{equation}
    \braket{\sigma_+} = \frac{1}{2}\bra{\phi} \mathcal{W}_1^{\dagger} \mathcal{W}_{2,p} \ket{\phi} \ .
\end{equation}\\

In this work, however, we will focus on Hamiltonians expressed in first quantization,
\begin{equation}
    \mathcal{H} = \frac{\bm{p}^2}{2m} + \mathcal{V}(\bm{r}) \ ,
    \label{1quantH}
\end{equation}
where $\bm{p}$ is the momentum, $m$ the mass, and $\mathcal{V}(\bm{r})$ the potential given as a function of the position $\bm{r}$. 
$\mathcal{H}$ describes an $M$-dimensional systems.
The time-evolution is directly performed in momentum and position space, discretized on a grid. 
The $N$ points, per dimension, of the grid are encoded in the basis states of $N_q=\log_2(N)$ qubits. 
The total number of qubits is then $MN_q$.
In this case, no explicit transformation to a basis representation is required, saving the numerical effort associated with the calculation of the integrals, which would scale as the square of the basis set size~\cite{manthe2002quantum}.

The expectation value of the Hamiltonian is now simply obtained from two sets of measurements, one in the momentum basis and one in the position basis.
This gives a clear advantage in the calculation of the right-hand side of Eq.~\ref{eq:vte}, which requires measuring $\bra{\psi(\bt)}\mathcal{H}\ket{\psi(\bt)}$ and $\Im(\bra{\partial_{\theta_k}\psi(\bt)}\mathcal{H}\ket{\psi(\bt)})$.
The first one is straightforward and reduces to measuring the expectation values $\bra{\psi(\bt)}\bm{p}^2\ket{\psi(\bt)}$ and $\bra{\psi(\bt)}\mathcal{V}(\bm{r})\ket{\psi(\bt)}$. 
In practice, this is obtained by repeatedly preparing and measuring the state $\ket{\psi(\bt)}$ in the computational (position) basis. 
For each outcome of the binary representation of $\bm{r}$, $\mathcal{V}(\bm{r})$ is classically computed and $\bra{\psi(\bt)}\mathcal{V}(\bm{r})\ket{\psi(\bt)}$ is obtained by averaging over all realizations of $\mathcal{V}(\bm{r})$. 
The same is done for the kinetic term by measuring in the momentum basis, \ie, applying a \gls{qft} right before the measurement (see Ref.~\cite{ollitrault2020non}). 

Let's now consider the calculation of the components $V_k$,
\begin{equation}
\begin{split}
    \Im\big(\bra{\partial_{\theta_k}\psi(\bt)}&\mathcal{H}\ket{\psi(\bt)}\big) = \\ &\frac{1}{2}\Re\big(\bra{\phi}\mathcal{W}(\bt)^{\dagger}\mathcal{H}\mathcal{U}(\bt)\ket{\phi}\big) \ ,
\end{split}
\end{equation}
where $\ket{\partial_{\theta_k}\psi(\bt)} = -\frac{i}{2} \mathcal{W}(\bt)\ket{\phi}$. 
In what follows, we simplify the notations of the quantum circuits as $\mathcal{U} = \mathcal{U}(\bt)$ and $\mathcal{W} = \mathcal{W}(\bt)$.

The scalar $\Re\big(\bra{\phi}\mathcal{W}^{\dagger}\mathcal{H}\mathcal{U}\ket{\phi}\big)$ is obtained in a similar way as in the second quantization case, namely with the circuit of Eq.~\eqref{eq:laflamme_circ} but by measuring both the ancilla qubit (in the computational basis) and the variational state register. This is given by the following quantum circuit,
\begin{equation*}
    \Qcircuit @C=0.5em @R=0.5em @!R {
	 	\lstick{  \ket{0} } &\gate{H} & \ctrl{1} & \gate{X} & \ctrl{1} & \gate{X} & \gate{H} & \meter\\
	 	\lstick{ \ket{\phi} } & \qw {/} & \gate{\mathcal{W}} & \qw & \gate{\mathcal{U}} & \qw & \qw  & \meter  }
\end{equation*}
We prove this by first considering the potential part of the Hamiltonian. In this case, $\bra{\phi} \mathcal{W}^{\dagger} \mathcal{V} \mathcal{U} \ket{\phi}$ is calculated from $\mathbb{E}[(-1)^s \mathcal{V}(j)]$ where $s$ is the measurement outcome of the ancilla qubit, while $j$ is that of the register. 
Indeed, the probability of measuring states $s$ and $j$ from the above circuit is
\begin{align}
    p(s,j) &= \big|\bra{s,j}\big(\ket{-}\bra{-}\otimes \mathcal{W} + \ket{+}\bra{+}\otimes \mathcal{U}\big)\ket{0,\phi}\big|^2 \nonumber\\
    &= \big|\braket{s|+}\braket{+|0}\otimes \bra{j}\mathcal{U}\ket{\phi} + \braket{s|-}\braket{-|0}\otimes\bra{j}\mathcal{W}\ket{\phi}\big|^2\nonumber\\
    &= \frac{1}{4} \big|\bra{j}\mathcal{U}\ket{\phi} + (-1)^s \bra{j}\mathcal{W}\ket{\phi}\big|^2 \nonumber\\
    &= \frac{1}{4} \bigg( |\bra{j}\mathcal{U}\ket{\phi}|^2 + |\bra{j}\mathcal{W}\ket{\phi}|^2  \nonumber\\ 
    & \qquad+ 2(-1)^s\Re\big(\bra{\phi}\mathcal{W}^{\dagger}\ket{j}\bra{j}\mathcal{U}\ket{\phi}\big) \bigg) \ .
\end{align}
With this, we finally obtain
\begin{align}
    \mathbb{E}[(-1)^s\mathcal{V}(j)] &= \sum_{s,j} (-1)^s\mathcal{V}(j)p(s,j) \nonumber\\
    & = \sum_j \mathcal{V}(j)\Re\big(\bra{\phi}\mathcal{W}^{\dagger}\ket{j}\bra{j}\mathcal{U}\ket{\phi}\big) \nonumber\\
    & = \Re(\bra{\phi}\mathcal{W}^{\dagger}\mathcal{V}\mathcal{U}\ket{\phi}) \ .
\end{align}
The same can be shown for the kinetic part of the Hamiltonian, provided the register is measured in the momentum basis, \ie, by introducing a \gls{qft} before the measurement.

\section{Heuristic variational forms}
\label{sec:heuristic}

\begin{figure}[h]
    \centering
    \includegraphics[width = \textwidth]{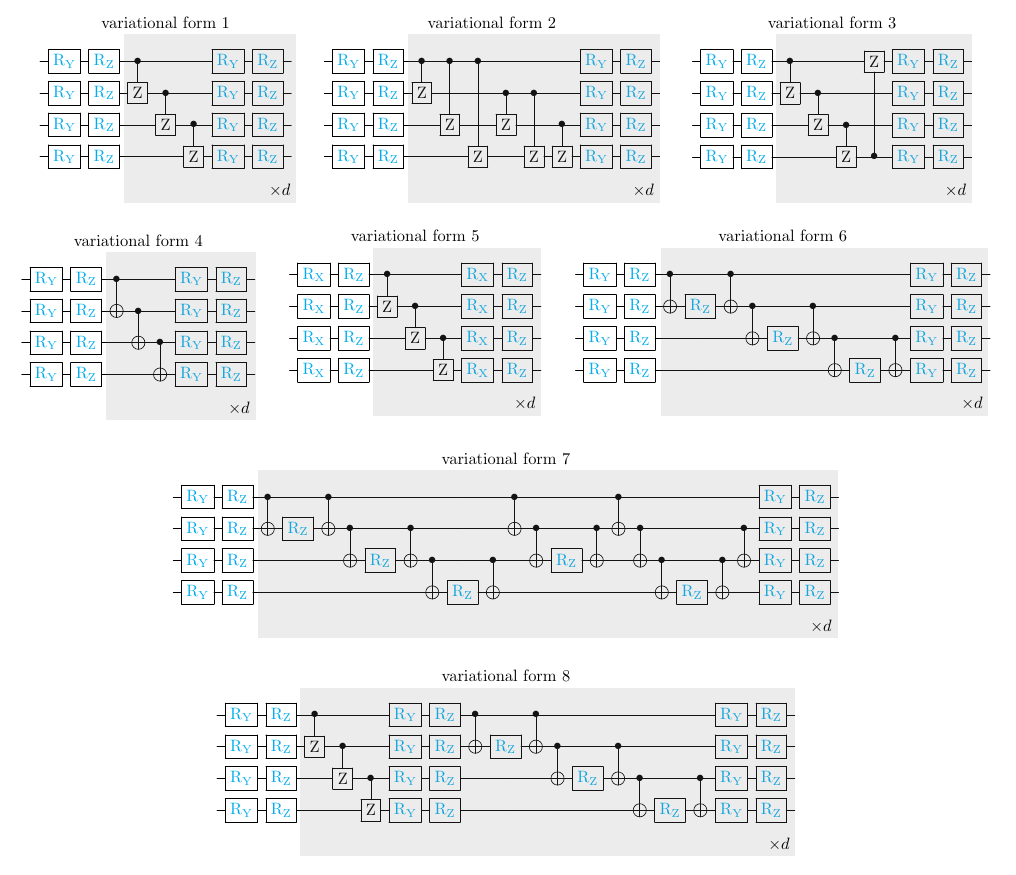}
    \caption{Quantum circuits for each variational form with depth $d$ on 4 qubits. The blue colour highlights a parameterized gate.}
    \label{fig:all_vfs_4q}
\end{figure}

The different variational forms employed in this work are summarized in Tab.~\ref{tab:all_vfs}. 
They all comprise alternating layers of single qubit rotations and entangling blocks. 
They differ in the choice of single qubit rotation gates, entangling gates, and coupling map, \ie, in the geometrical way the qubits are coupled to each other. In the linear coupling map, each qubit is coupled to its two nearest neighbors only. The circular map is similar but adds a coupling between the first and last qubits. Finally, in the full coupling map, each qubit is coupled to all other qubits. 
For the sake of clarity, Fig.~\ref{fig:all_vfs_4q} gives concrete examples of the resulting circuits for each variational form for four qubits.
\begin{table}[h]
\centering
\begin{tabular}{|c|c|c|c|c|}
\hline
\hline
     Name & Single qubit gates & Entangling gates & Coupling map & \thead{\# params. for $N_q$ \\qubits and depth $d$} \\
     \hline
     \hline
     vf1 & \makecell{\Qcircuit @C=0.5em @R=0.5em @!R {
    \lstick{} & \gate{\ry} & \gate{\rz} & \qw}} & \makecell{\Qcircuit @C=0.5em @R=0.5em @!R {
    \lstick{} & \ctrl{1} & \qw \\ \lstick{} & \gate{\z} & \qw} \vspace{0.1cm}}& linear & $2N_q(d+1)$ \\
    \hline
     vf2 & \makecell{\Qcircuit @C=0.5em @R=0.5em @!R {
    \lstick{} & \gate{\ry} & \gate{\rz} & \qw}} & \makecell{\Qcircuit @C=0.5em @R=0.5em @!R {
    \lstick{} & \ctrl{1} & \qw \\ \lstick{} & \gate{\z} & \qw} \vspace{0.1cm}}& full & $2N_q(d+1)$ \\ 
    \hline
     vf3 & \makecell{\Qcircuit @C=0.5em @R=0.5em @!R {
    \lstick{} & \gate{\ry} & \gate{\rz} & \qw}} & \makecell{\Qcircuit @C=0.5em @R=0.5em @!R {
    \lstick{} & \ctrl{1} & \qw \\ \lstick{} & \gate{\z} & \qw} \vspace{0.1cm}}& circular & $2N_q(d+1)$ \\ 
    \hline
     vf4 & \makecell{\Qcircuit @C=0.5em @R=0.5em @!R {
    \lstick{} & \gate{\ry} & \gate{\rz} & \qw}} & \makecell{\Qcircuit @C=0.5em @R=0.5em @!R {
    \lstick{} & \ctrl{1} & \qw \\ \lstick{} & \targ & \qw} \vspace{0.1cm}}& linear & $2N_q(d+1)$ \\ 
    \hline
     vf5 & \makecell{\Qcircuit @C=0.5em @R=0.5em @!R {
    \lstick{} & \gate{\rx} & \gate{\rz} & \qw}} & \makecell{\Qcircuit @C=0.5em @R=0.5em @!R {
    \lstick{} & \ctrl{1} & \qw \\ \lstick{} & \gate{\z} & \qw} \vspace{0.1cm}}& linear & $2N_q(d+1)$ \\ 
    \hline
    vf6 & \makecell{\Qcircuit @C=0.5em @R=0.5em @!R {
    \lstick{} & \gate{\ry} & \gate{\rz} & \qw}} & \makecell{\Qcircuit @C=0.5em @R=0.5em @!R {
    \lstick{} & \ctrl{1} & \qw & \ctrl{1} & \qw\\ \lstick{} & \targ & \gate{\rz} & \targ & \qw} \vspace{0.1cm}}& linear & \thead{$2N_q(d+1)$\\$ + d(N_q-1)$} \\ 
    \hline
    vf7 & \makecell{\Qcircuit @C=0.5em @R=0.5em @!R {
    \lstick{} & \gate{\ry} & \gate{\rz} & \qw}} & \makecell{\Qcircuit @C=0.5em @R=0.5em @!R {
    \lstick{} & \ctrl{1} & \qw & \ctrl{1} & \qw\\ \lstick{} & \targ & \gate{\rz} & \targ & \qw } \hspace{0.1cm} \Qcircuit @C=0.5em @R=0.5em @!R {
    \lstick{} & \ctrl{1} & \qw & \qw & \qw & \ctrl{1} & \qw\\ \lstick{} & \targ & \ctrl{1} & \qw & \ctrl{1} & \targ & \qw \\
    \lstick{} & \qw & \targ & \gate{\rz} & \targ & \qw & \qw} \vspace{0.1cm}}& linear & \thead{$2N_q(d+1) $\\$+ d(N_q-1)$ \\$ + d(N_q-2)$} \\ 
    \hline
    vf8 & \makecell{\Qcircuit @C=0.5em @R=0.5em @!R {
    \lstick{} & \gate{\ry} & \gate{\rz} & \qw}} & \makecell{\Qcircuit @C=0.5em @R=0.5em @!R { \lstick{} & \ctrl{1} & \qw \\ \lstick{} & \gate{\z} & \qw } \hspace{0.1cm} \Qcircuit @C=0.5em @R=0.5em @!R {\lstick{} & \ctrl{1} & \qw & \ctrl{1} & \qw\\ \lstick{} & \targ & \gate{\rz} & \targ & \qw }\hspace{0.1cm}   \vspace{0.1cm}}& linear & \thead{$2N_q(2d+1)$\\$ + d(N_q-1) $} \\ 
    \hline
\end{tabular}
\caption{Definitions of the different variational forms employed throughout this work. The blue colour highlights a parameterized gate.}
\label{tab:all_vfs}
\end{table} 

\section{Comparing variational forms}
\label{section:comparision_vfs}

\begin{figure}[h!]
    \centering
    \includegraphics[width=0.8\textwidth]{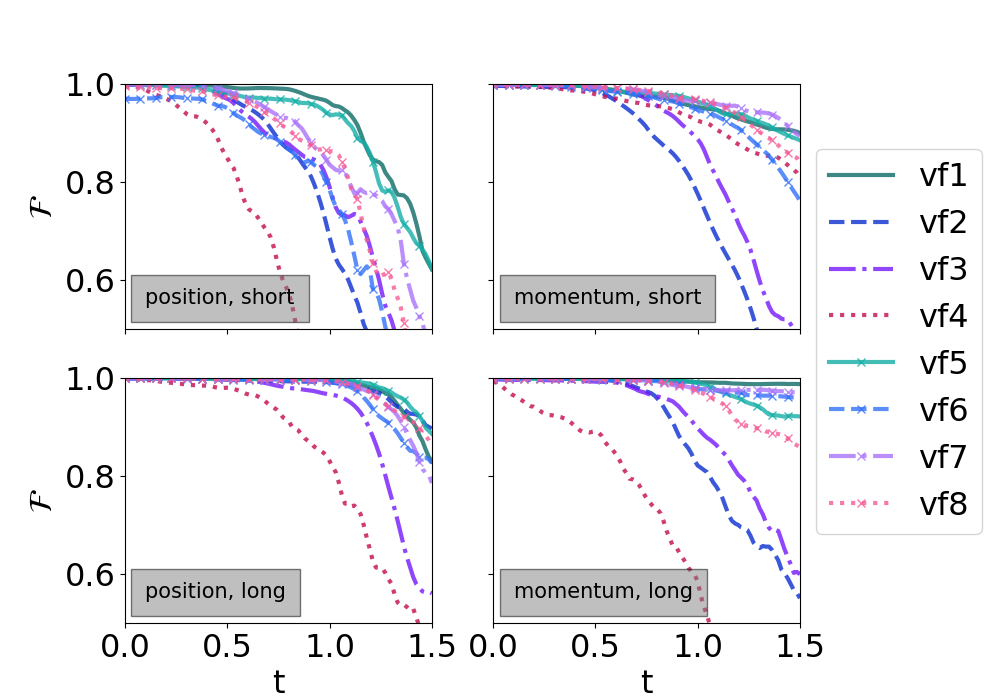}
    \caption{Fidelity of the variationally time-evolved wavefunction with respect to the exact one as a function of time for \textit{top-left} all variational forms with short depth in position space, \textit{top-right} all variational forms with short depth in momentum space, \textit{bottom-left} all variational forms with large depth in position space and \textit{bottom-left} all variational forms with large depth in momentum space.
    All varioational forms are defined in Section~\ref{sec:heuristic}.
    The specific depths and corresponding number of parameters are given in Tab.~\ref{tab:vf_explicit_numbers} together. These results are obtained for the case of the Eckart barrier with 6 qubits.}
    \label{fig:fid_all_vfs}
\end{figure}

One of the main advantages of the variational approach described in this work for first quantized Hamiltonians is the possibility to perform the dynamics without explicitly including the Hamiltonian in the quantum circuit. Instead, heuristic variational forms are exploited.
There exist many ways the heuristic variational forms can be defined. 
Here, we test several ones for the dynamics of the nuclear wavepacket colliding with an Eckart barrier (6 qubits) and compare the resulting fidelities over the simulation time. All variational forms are defined in section \ref{sec:heuristic} of this \si{} and we explicitly give the depths and corresponding number of parameters in Tab.~\ref{tab:vf_explicit_numbers}.
\begin{table}[h!]
    \centering
    \begin{tabular}{|c|c|c|c|c|}
        \hline
        \hline
         Variational forms & Depth short & \# params short & Depth long & \# params long \\
         \hline
         \hline
         vf1 & 5 & 72 & 8 & 108\\ 
         \hline
         vf2 & 5 & 72 & 8 & 108\\ 
         \hline
         vf3 & 5 & 72 & 8 & 108\\ 
         \hline
         vf4 & 5 & 72 & 8 & 108\\ 
         \hline
         vf5 & 5 & 72 & 8 & 108\\ 
         \hline
         vf6 & 3 & 63 & 5 & 97\\ 
         \hline
         vf7 & 3 & 75 & 4 & 96\\ 
         \hline
         vf8 & 2 & 70 & 3 & 99\\ 
         \hline
    \end{tabular}
    \caption{Depth and number of parameters for each of the variational forms tested on the simulation of the dynamics of a wavepacket colliding with an Eckart barrier using 6 qubits. Their performance is shown on Fig.~\ref{fig:fid_all_vfs}}
    \label{tab:vf_explicit_numbers}
\end{table}
From the results shown in Fig.~\ref{fig:fid_all_vfs} one can see that, in general, an increase in accuracy is obtained by enlarging the depth and by going to momentum space. 
However, no clear trends can be identified as to how to construct the variational form. Nonetheless, variational forms 1 and 5 (with linear connectivity of controlled-Z entangling gates), seem to generally perform the best. 

\section{Time evolution of the variational parameters}
\begin{figure*}[h]
     \centering
     \begin{subfigure}[b]{0.32\textwidth}
         \centering
         \includegraphics[width=\textwidth]{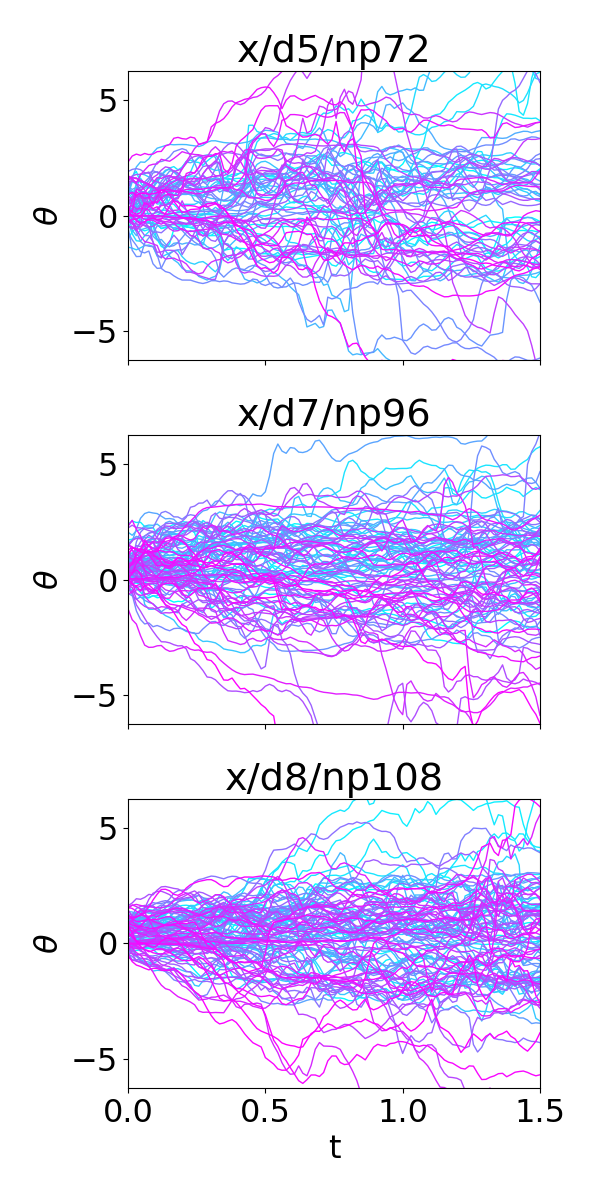}
         \caption{Free particle}
         \label{fig:FP_xparams}
     \end{subfigure}
     \begin{subfigure}[b]{0.32\textwidth}
         \centering
         \includegraphics[width=\textwidth]{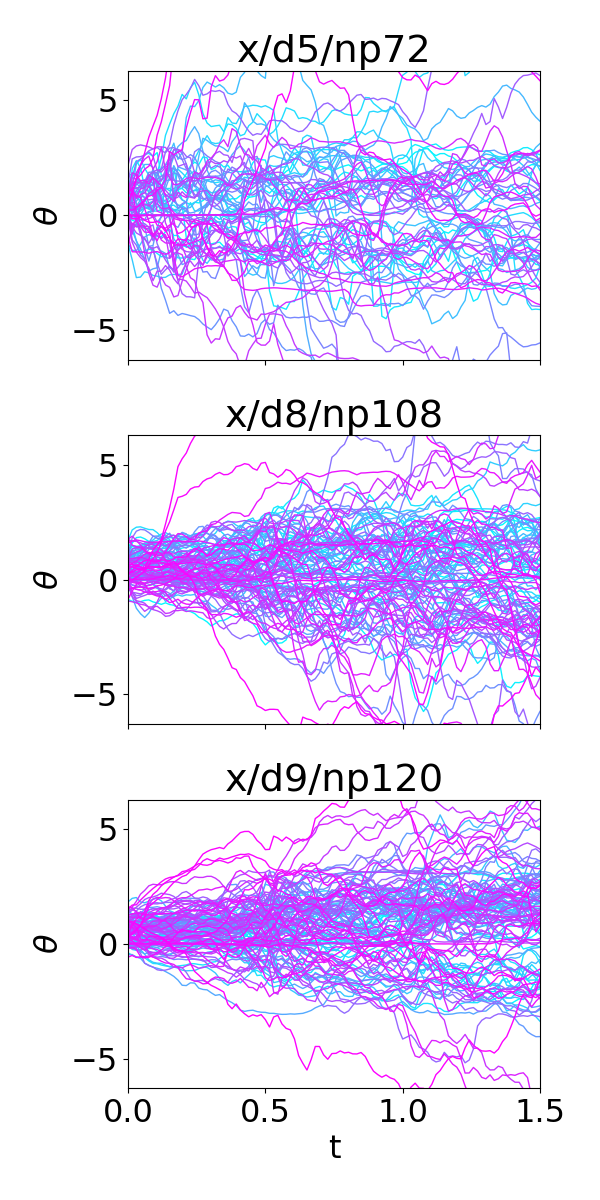}
         \caption{Harmonic oscillator}
         \label{fig:HO_xparams}
     \end{subfigure}
     \begin{subfigure}[b]{0.32\textwidth}
         \centering
         \includegraphics[width=\textwidth]{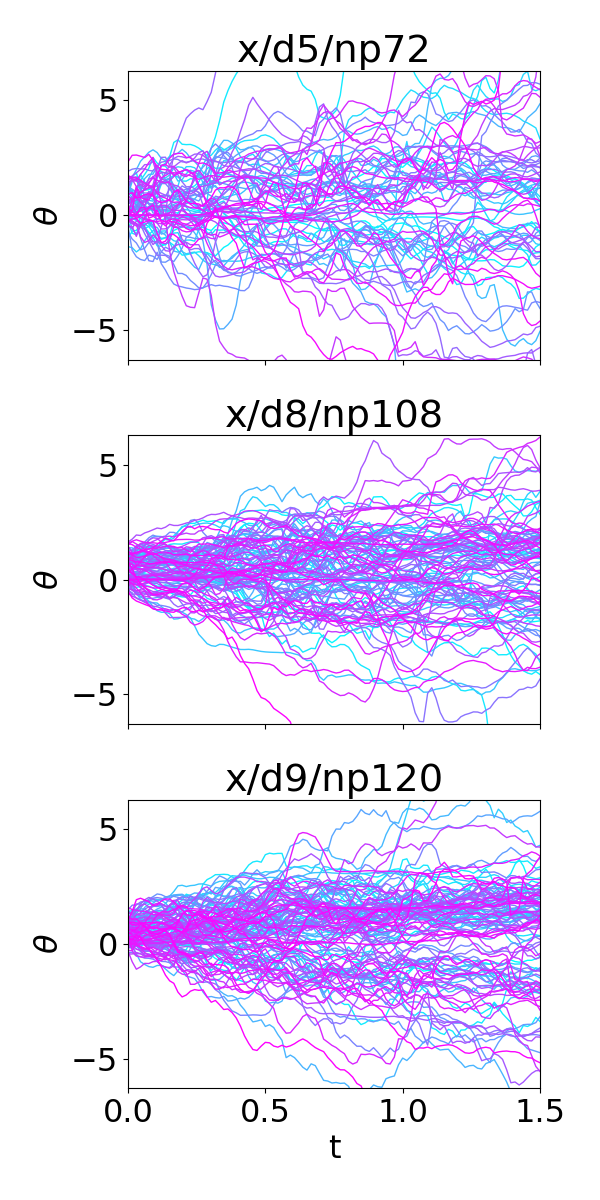}
         \caption{Eckart barrier}
         \label{fig:EB_xparams}
     \end{subfigure}
        \caption{Time evolution of the variational parameters for each one-dimensional system obtained with variational form vf1 in position space and at given depth, $d$ (with $n_p$ variational parameters).}
        \label{fig:params_evolution_xspace}
\end{figure*}

\begin{figure*}[h]
     \centering
     \begin{subfigure}[b]{0.32\textwidth}
         \centering
         \includegraphics[width=\textwidth]{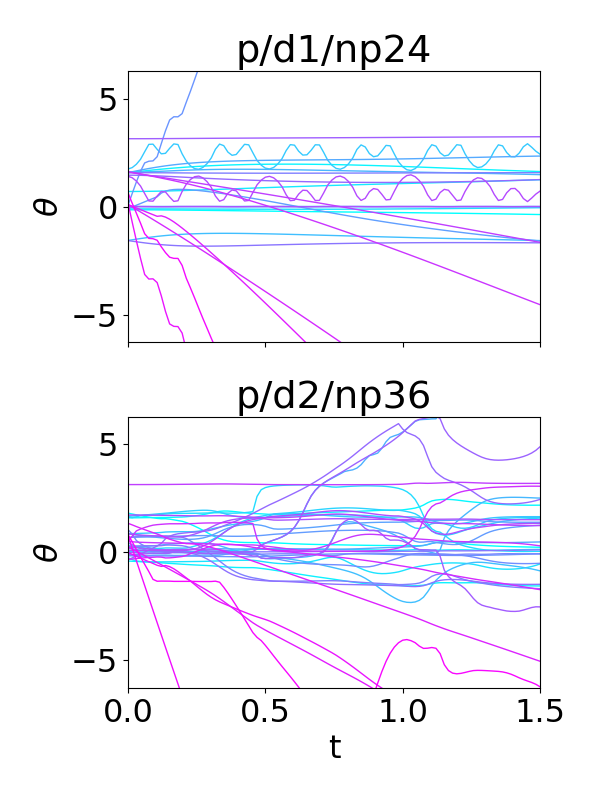}
         \caption{Free particle}
         \label{fig:FP_pparams}
     \end{subfigure}
     \begin{subfigure}[b]{0.32\textwidth}
         \centering
         \includegraphics[width=\textwidth]{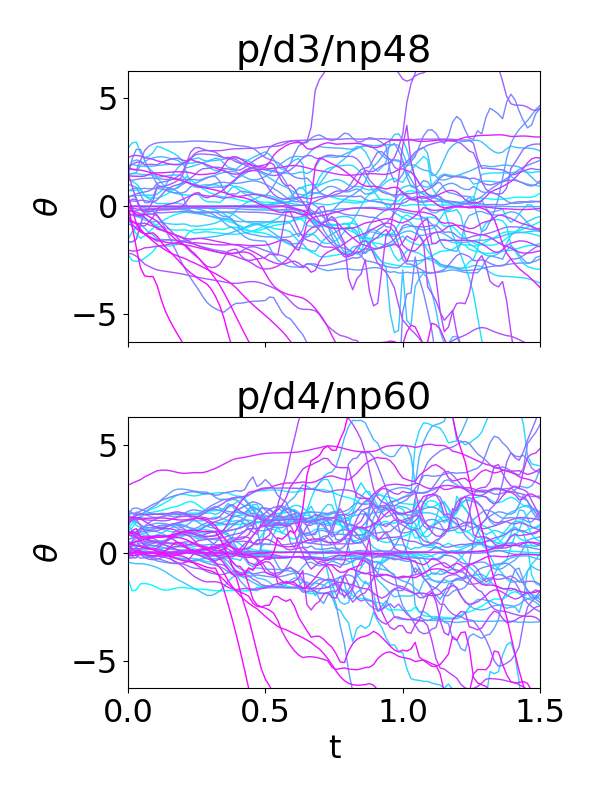}
         \caption{Harmonic oscillator}
         \label{fig:HO_pparams}
     \end{subfigure}
     \begin{subfigure}[b]{0.32\textwidth}
         \centering
         \includegraphics[width=\textwidth]{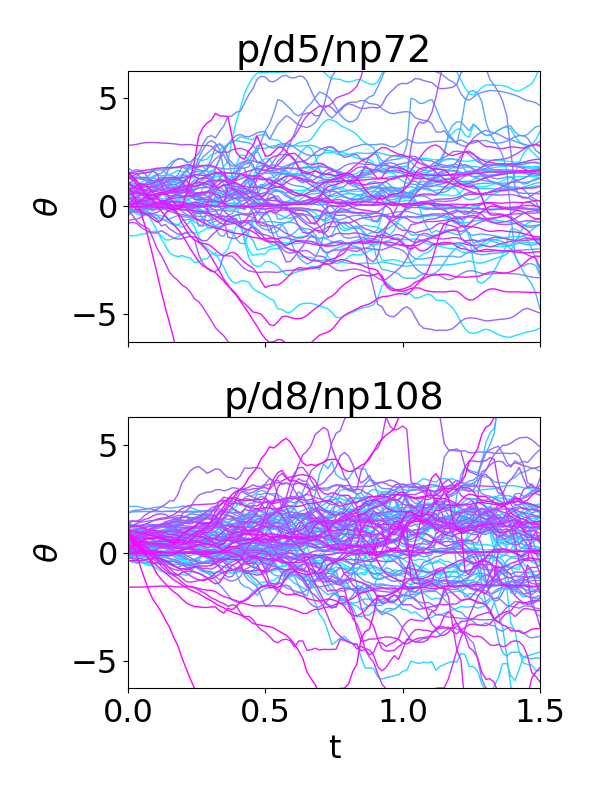}
         \caption{Eckart barrier}
         \label{fig:EB_pparams}
     \end{subfigure}
        \caption{Time evolution of the variational parameters for each one-dimensional system obtained with variational form vf1 in momentum space and at given depth, $d$ (with $n_p$ variational parameters).}
        \label{fig:params_evolution_pspace}
\end{figure*}

\begin{figure*}[h]
     \centering
     \begin{subfigure}[b]{0.32\textwidth}
         \centering
         \includegraphics[width=\textwidth]{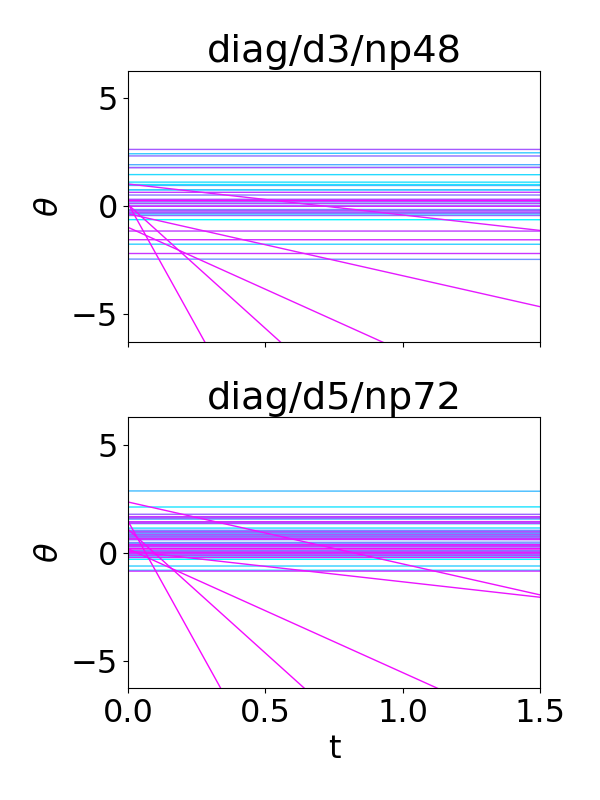}
         \caption{Harmonic oscillator}
         \label{fig:FP_dparams}
     \end{subfigure}
     \begin{subfigure}[b]{0.32\textwidth}
         \centering
         \includegraphics[width=\textwidth]{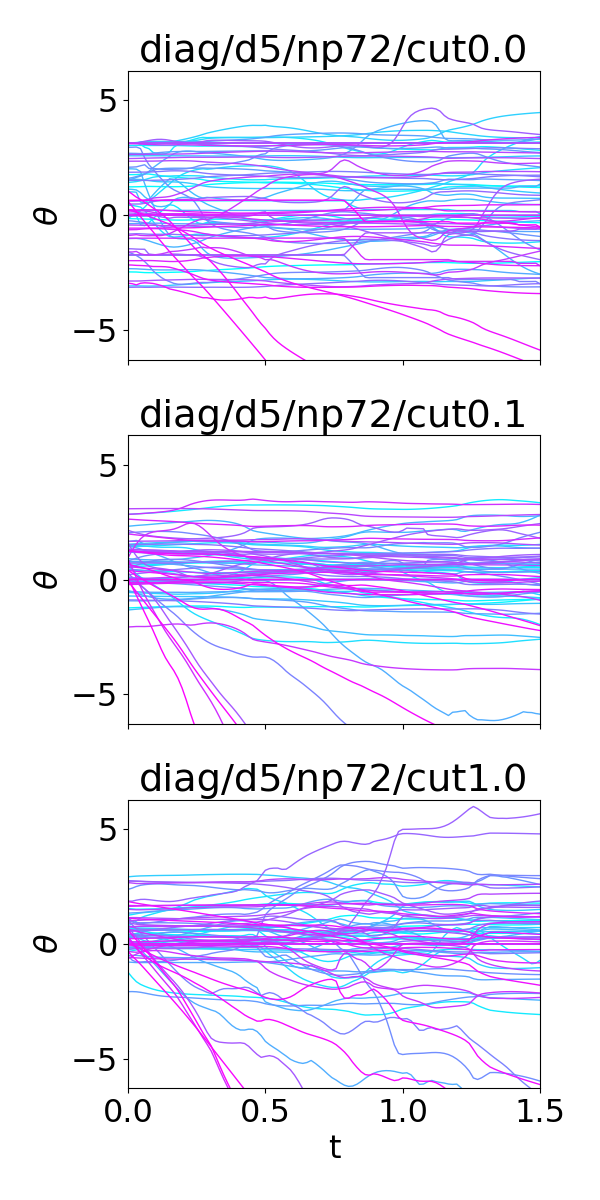}
         \caption{Eckart barrier}
         \label{fig:HO_dparams}
     \end{subfigure}
        \caption{Time evolution of the variational parameters for two of the one-dimensional systems obtained with variational form vf2 in diagonal space and at given depth, $d$ (with $n_p$ variational parameters), and cutoff value $cut$.}
        \label{fig:params_evolution_dspace}
\end{figure*}

In this section, we show the dynamics of the parameters obtained with the different \glspl{vte} to supplement the fidelity results shown in the main text.
In Fig.~\ref{fig:params_evolution_xspace}, we report the evolution in position space, in Fig.~\ref{fig:params_evolution_pspace} those in momentum space, and, finally in Fig.~\ref{fig:params_evolution_dspace}, the evolutions in diagonal space.
The first two cases (position and momentum) are obtained with variational form vf1, while in the last case (diagonal), we use variational form vf2.

For all systems, in position space, we observe sharp changes in the dynamics. In general, the parameter values also diverge with time.  
The same observations can be made in momentum space for the harmonic oscillator and the Eckart barrier. 

Whenever the circuit diagonalizes the Hamiltonian as in Fig.~\ref{fig:params_evolution_pspace}(a) and Fig.~\ref{fig:params_evolution_dspace}(a) and (b), the parameter evolutions become smooth or even trivial. 
This is unsurprising as, in this case, there is no transfer of amplitudes but only evolution of the phases. 
Note that in Fig.~\ref{fig:params_evolution_dspace}(b), when $cut>0$, the partial diagonalization of the Hamiltonian suffices to ensure a smoother evolution of the parameters. 

\section{Study on the accuracy of the time evolution in one dimension}
\label{sec:accuracy}
To better understand the origin of the errors occurring during the \gls{vte}, we perform additional numerical simulations. Unless explicitly stated otherwise, the results of this section are obtained with variational forms in position space.

\begin{figure}[h]
    \centering
    \includegraphics[width=0.35\textwidth]{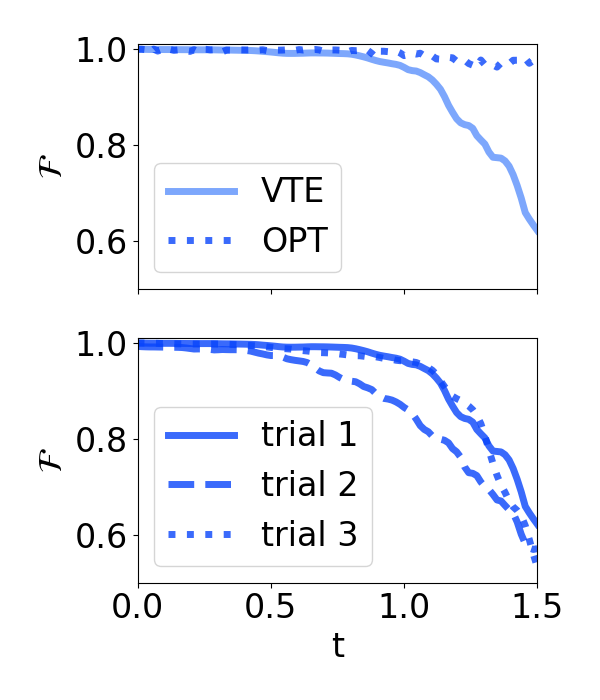}
    \caption{\textit{Top:} Fidelity of the variationally time-evolved wavefunction with respect to the exact one as a function of time (full line) and after being optimized at each time step to maximize the fidelity (dotted line).\\ 
    \textit{Bottom:} Fidelity of the variationally time-evolved wavefunction with respect to the exact one as a function of time for three different sets of initial parameters. \\
    These results are obtained for the case of the Eckart barrier with 6 qubits and variational form vf1, depth 5.}
    \label{fig:opt_and_trials}
\end{figure}

We start with the Eckart barrier case. 
The first step is to identify whether the variational form is flexible enough to represent the state at each time step. 
For this, we optimize the parameters obtained from \gls{vte} in 6 qubits with depth 5 and variational form vf1 to maximize $\mathcal{F}(t)$ at each time step (see definition in main text).
The results displayed in the top panel of Fig.~\ref{fig:opt_and_trials} show that we can indeed get a better fidelity for times $t>0.8$. 
This suggests that the variational form is good enough to represent, with good accuracy, the exact wavefunction through the entire dynamics.

There exist several degenerate sets of variational parameters which represent the same wavefunction. In other words, the initial state can be prepared with a given accuracy from different sets of parameters using the same \ansatz{}. In the bottom panel of Fig.~\ref{fig:opt_and_trials}, we show how the accuracy of the dynamics can be affected by the choice these initial parameters. 
For the three different trials displayed in Fig.~\ref{fig:opt_and_trials}, we find those initial parameters by maximizing $\mathcal{F}(t=0)$. Each one of these three optimization processes starts with random guesses and converges to above $99\%$ fidelity to different parameters solutions.
Fig.~\ref{fig:opt_and_trials} clearly shows a difference in the fidelities of the dynamics for the three trials.
However, we observe that choosing the best set of initial parameters is non-trivial and cannot be made by simply looking at the initial fidelity or early time observables such as the local-in-time error~\cite{martinazzo2020}. \\

As a second step, we aim to study the effect of the different numerical parameters on the accuracy.
We choose to work with the simplest system: a free particle. 
The space is discretized with 5 qubits. 
The initial conditions are $(x_0,p_0) = (0, 5)$. 
The width of the initial wavepacket is set to $\mathcal{B}=1/\sqrt{2}$. 
The parametrized circuit corresponds to variational form vf2 with depth 3. 

\begin{figure}[h]
    \centering
    \includegraphics[width=0.4\textwidth]{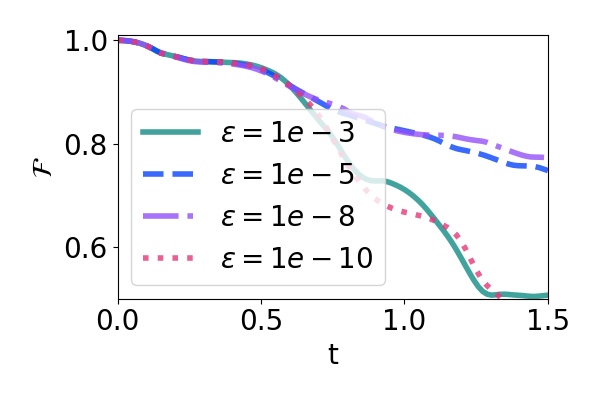}
    \caption{Fidelity of the variationally time-evolved wavefunction with respect to the exact one as a function of time and for different values of the finite difference step, $\epsilon$. These results are obtained for the case of the free particle with 5 qubits and variational form vf2, depth 3.}
    \label{fig:finitediffs}
\end{figure}

We first run the \gls{vte} for different values of the finite difference step size, $\epsilon$. The results are displayed in Fig.~\ref{fig:finitediffs}, showing identical fidelities in the relevant part of the evolution when the accuracy is above 95\%. 

\begin{figure}[h]
    \centering
    \includegraphics[width=0.4\textwidth]{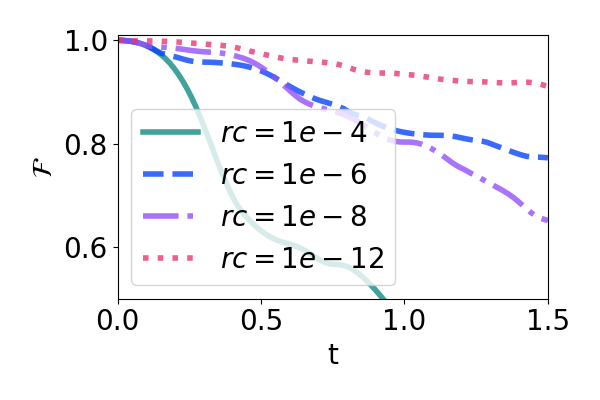}
    \caption{Fidelity of the variationally time-evolved wavefunction with respect to the exact one as a function of time. Different ratios, $rc$, for cutting off small singular values are employed. These results are obtained for the case of the free particle with 5 qubits and variational form vf2, depth 3.}
    \label{fig:rcond}
\end{figure}

We then fix $\epsilon$ back to its original value of $10^{-8}$ and change the reconditioning number, $rc$, the ratio for cutting off small singular values in the least-squares algorithm. 
The results displayed in Fig.~\ref{fig:rcond}(a) show that the fidelity increases with smaller $rc$. 
This implies numerical errors coming from instabilities in the inversion of the matrix F when solving Eq.~\ref{eq:vte} (main text).

\begin{figure}[h]
    \centering
    \includegraphics[width=0.4\textwidth]{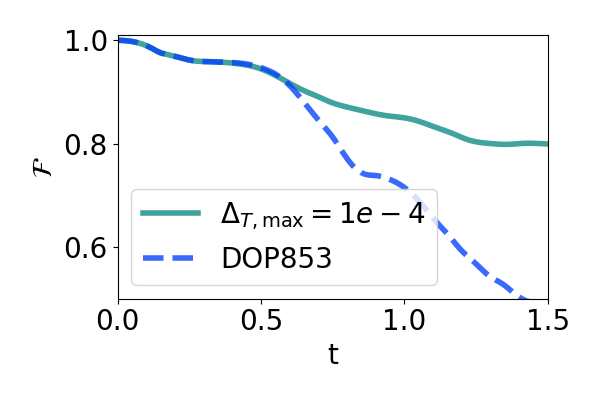}
    \caption{Fidelity of the variationally time-evolved wavefunction with respect to the exact one as a function of time, obtained with different ordinary differential equation solvers. The full line corresponds to the Runge-Kutta 5(4) solver used throughout this work but with a maximal time step fixed at $10^{-4}$. On the other hand, the dashed line was obtained with an explicit Runge-Kutta method of order 8 as implemented in \textsc{SciPy}~\cite{2020SciPy-NMeth}. These results are obtained for the case of the free particle with 5 qubits and variational form vf2, depth 3.}
    \label{fig:solvers}
\end{figure}

We then study the performance of the ordinary differential equation solver. 
We employ the same Runge-Kutta 5(4) solver, but now fix the maximum time step to $10^{-4}$, and compare to an explicit Runge-Kutta solver of order 8 (DOP853)~\cite{2020SciPy-NMeth}.
As shown in Fig.~\ref{fig:solvers}, the results of these two simulations are identical in the first part of the dynamics (before the accuracy drops below 90\%) and differ afterwards.

\begin{figure}[h]
    \centering
    \includegraphics[width=0.4\textwidth]{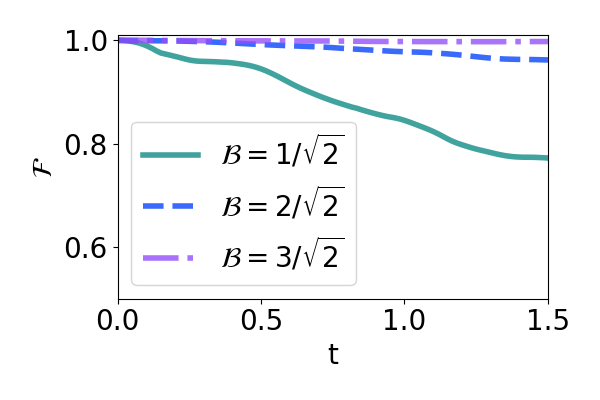}
    \caption{Fidelity of the variationally time-evolved wavefunction with respect to the exact one as a function of time and for different values of the width, $\mathcal{B}$, of the initial wavepacket. These results are obtained for the case of the free particle with 5 qubits and variational form vf2, depth 3.}
    \label{fig:sigmas}
\end{figure}

Finally, we vary the initial width $\mathcal{B}$ of the wavepacket. Interestingly, we observe that the results are much improved when increasing the initial width (see Fig.~\ref{fig:sigmas}). 
The latter influences the overall spread of the wavepacket during the evolution. The difference between initial and final width is $0.841$, $0.148$ and $0.028$ when $\mathcal{B}=1/\sqrt{2}$, $\mathcal{B}=2/\sqrt{2}$ and $\mathcal{B}=3/\sqrt{2}$, respectively.
These results suggest that the spread of the wavepacket is difficult to capture with the grid-based \gls{vte}. \\

\begin{figure}[h]
    \centering
    \includegraphics[width=0.35\textwidth]{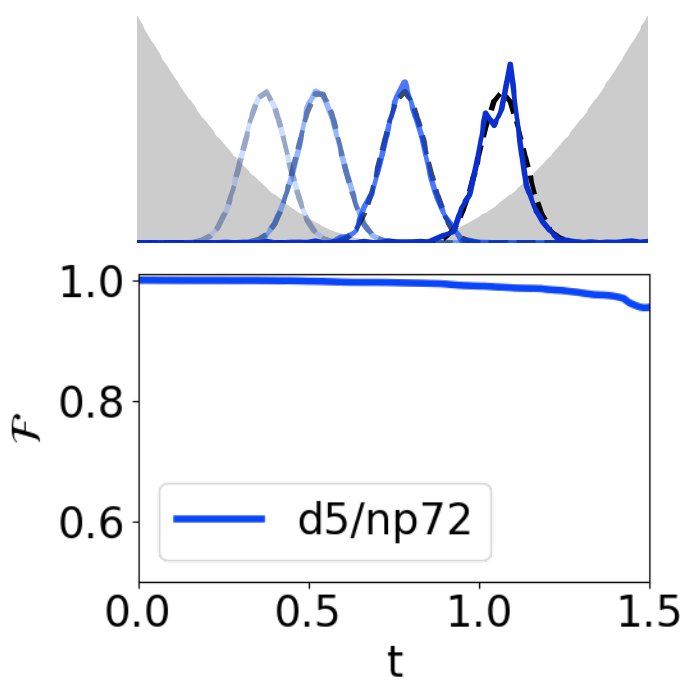}
    \caption{\textit{Top} Snapshots of the exact (dashed lines) and the variationally time-evolved (full lines) modulus squared wavefunctions at times $t=0.00, 0.45, 0.91$, and $1.5$ (lightest to darkest curve, respectively) of the harmonic oscillator with a non-spreading width $\mathcal{B}=0.6$.\\
    \textit{Bottom} Fidelity of the variationally time-evolved wavefunction with respect to the exact one as a function of time. These results are obtained for the case of the harmonic oscillator with width $\mathcal{B}=0.6$, 6 qubits, and variational form vf1 depth 5 in position space.}
    \label{fig:no_spread}
\end{figure}

To validate those observations, we run the \gls{vte} of a wavepacket oscillating in a harmonic potential without spread. 
The space is discretized in 6 qubits as in the simulations of the main text. 
We also keep the same initial conditions $(x_0,p_0) = (-3.5, 2)$, and the same variational form, vf1 depth 5 in position space.
This time, however, the initial width is taken to be $\mathcal{B}=\mathcal{B}_{\text{gs}}=0.6$, where $\mathcal{B}_{\text{gs}}$ is the width of Hamiltonian's ground state. 
Fig.~\ref{fig:no_spread} confirms our previous observations: the results in the case of a non-changing width are much improved compared to the results of the main text in which the wavepacket's width changes over time. 
This leads to the conclusion that the \gls{vte} performs reasonably well (with few parameters) for simple dynamics.\\

\begin{figure*}[h]
     \centering
     \begin{subfigure}[b]{0.32\textwidth}
         \centering
         \includegraphics[width=\textwidth]{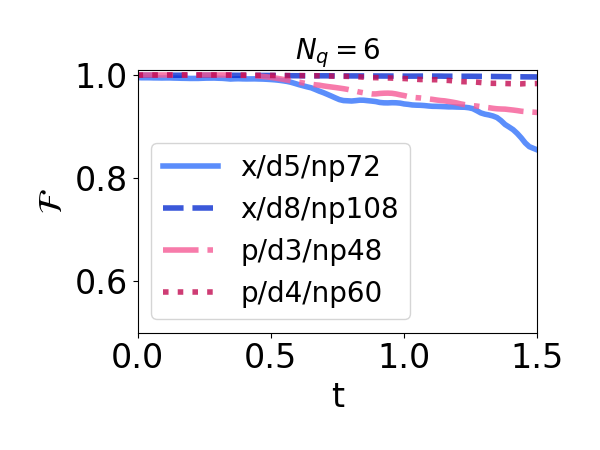}
         \caption{64 grid points}
         \label{fig:HO_nq6}
     \end{subfigure}
     \begin{subfigure}[b]{0.32\textwidth}
         \centering
         \includegraphics[width=\textwidth]{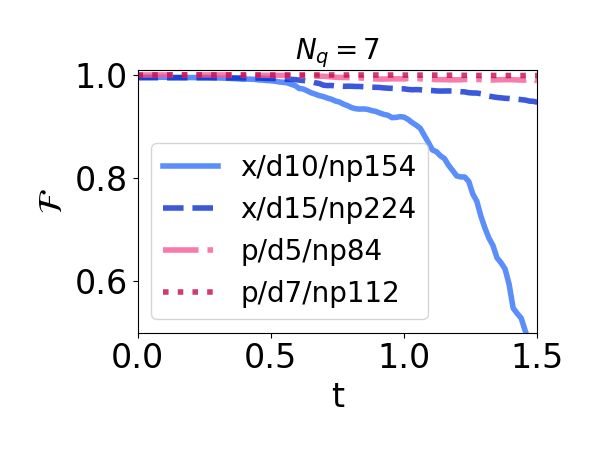}
         \caption{128 grid points}
         \label{fig:HO_nq7}
     \end{subfigure}
     \begin{subfigure}[b]{0.32\textwidth}
         \centering
         \includegraphics[width=\textwidth]{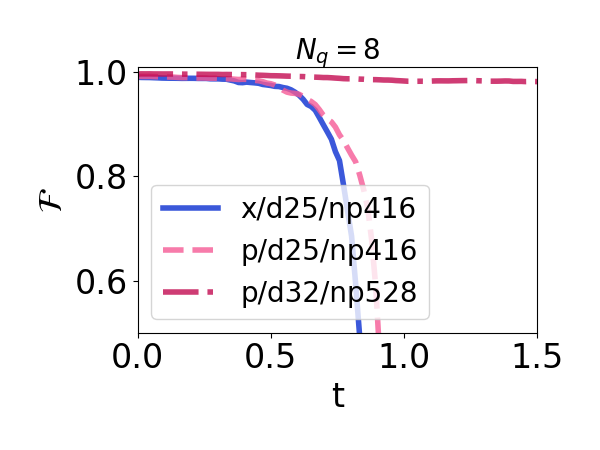}
         \caption{256 grid points}
         \label{fig:HO_nq8}
     \end{subfigure}
        \caption{Fidelity of the variationally time-evolved wavefunction with respect to the exact one as a function of time. These results are obtained for the case of the harmonic oscillator with variational form vf1. The space is discretized with (a) 6 qubits, (b) 7 qubits, and (c) 8 qubits.}
        \label{fig:HO_nq}
\end{figure*}

We then study the effect of increasing the precision in the grid mesh on the dynamics of the harmonic oscillator. The initial conditions are the ones of the main text: $\mathcal{B}=1/\sqrt{2}$ and $(x_0,p_0) = (-3.5, 2)$. The space is discretized with 6, 7 and 8 qubits corresponding to 64, 128 and 256 grid points, respectively. The variational form is always taken to be vf1. The results are shown in Fig.~\ref{fig:HO_nq}. We employ both the position and momentum representations of the wavefunction (indicated with x and p in Fig.~\ref{fig:HO_nq}, respectively). The different depths and corresponding number of parameters are also indicated in the legend.
The fidelities shown in Fig.~\ref{fig:HO_nq} are computed from the variationally time evolved wavefunctions with respect to the exactly evolved ones discretized on the same grid. In other words, both the reference and the variational wavefunctions are expressed in the same number of qubits. 
In all cases, we see that the correct dynamics are recovered in the limit of the number of parameters, $n_p$, approaching the size of the Hilbert space. 
In the momentum space, the number of parameters needed to maintain an accuracy of $\mathcal{F}>95\%$ over the whole time range remains low for $N_q = 6$ and $N_q = 7$. It also shows a good scaling behavior when going from 6 to 7 qubits. 
Indeed, if 60 parameters are necessary in the 6-qubit case, this number only raises to 84 when we double the number of grid points (7 qubits). 
However, we observe a drop in accuracy when going to 8 qubits. In this case, 416 parameters (for a full Hilbert space represented with 510 real parameters) were not enough to get accurate dynamics both in position and momentum spaces. 
These results show the influence of the grid mesh on the accuracy of the \gls{vte}. The observed non-monotonic behavior pinpoints the strong correlation existing between the different numerical factors, such as $N_q$ and $rc$.
It is important to note that those numerical instabilities are always corrected for when the number of parameters is high enough. 

\begin{figure*}[h]
     \centering
     \begin{subfigure}[b]{0.32\textwidth}
         \centering
         \includegraphics[width=\textwidth]{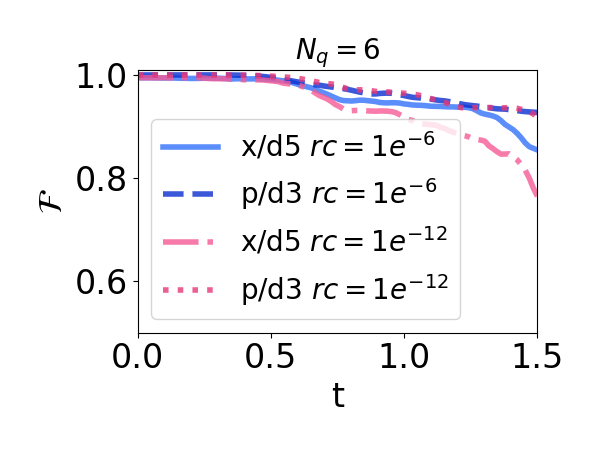}
         \caption{64 grid points}
         \label{fig:HO_rc_nq6}
     \end{subfigure}
     \begin{subfigure}[b]{0.32\textwidth}
         \centering
         \includegraphics[width=\textwidth]{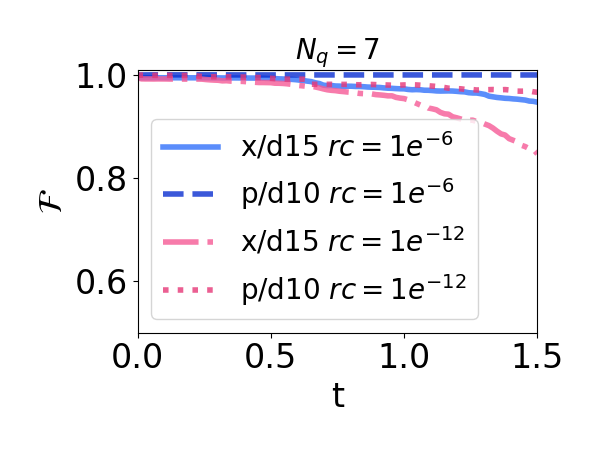}
         \caption{128 grid points}
         \label{fig:HO_rc_nq7}
     \end{subfigure}
    \begin{subfigure}[b]{0.32\textwidth}
         \centering
         \includegraphics[width=\textwidth]{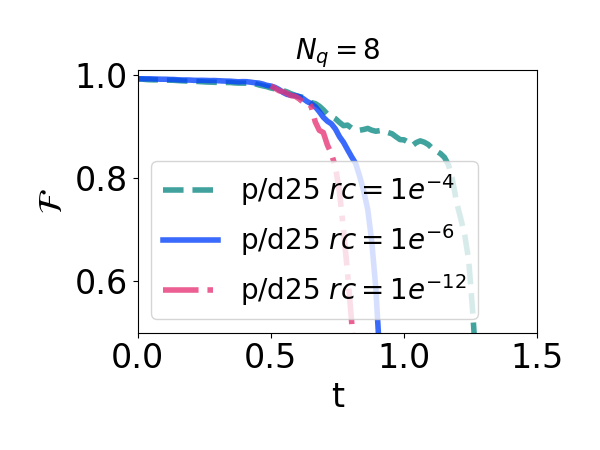}
         \caption{256 grid points}
         \label{fig:HO_rc_nq8}
     \end{subfigure}
        \caption{Fidelity of the variationally time-evolved wavefunction with respect to the exact one as a function of time. These results are obtained for the case of the harmonic oscillator with variational form vf1, and for different values $rc$ for cutting off small singular values. The space is discretized with (a) 6 qubits, (b) 7 qubits, and (c) 8 qubits.}
        \label{fig:HO_rc_nq}
\end{figure*}

We repeat the simulation of the same harmonic oscillator system for different values of $rc$, the cutoff ratio of small singular value for the inversion of the matrix F.
Those dynamics are obtained with 6, 7, and 8 qubits. The results are shown in Fig.~\ref{fig:HO_rc_nq}. We employ variational form vf1 in position (x) and momentum (p) space, and for different depths (d) as indicated in the legend of Fig.~\ref{fig:HO_rc_nq}.
As opposed to the results of Fig.~\ref{fig:rcond}, in this case, decreasing the value of $rc$ does not improve the results but even worsens them slightly. 
On the other hand, we observe improved results when increasing $rc$ with thinner a grid mesh (Fig.~\ref{fig:HO_rc_nq}(c)).
This shows again that the numerical effects are correlated and system dependent. \\

In conclusion, we highlight the following point:
\begin{itemize}
    \item Heuristic variational forms have the flexibility to accurately and efficiently, \ie, with few variational parameters, describe the targeted wavefunctions at all times of their dynamics. The loss of accuracy observed throughout the different simulations is an inherent effect of the method.
    \item When the number of variational parameters is insufficient, the dynamics strongly depend on the numerical setup of the simulation (grid mesh size, reconditioning number, initial parameters, etc). 
    \item The correct dynamics are always recovered by increasing the number of variational parameters. In this case, the algorithm is stable. 
    \item The more complex the dynamics is, the larger is the number of needed parameters. The term \textit{complex} relates to the size of the energy eigenspace involved in the dynamics (number of eigenvectors of $\mathcal{H}$ spanning the subspace in which the dynamics is defined). In practice, few symmetries and a large range of positions/frequencies involved are characteristics of complex dynamics when working with the canonical first quantized representation of the Hamiltonian.
    \item In all cases presented here, the expression of the wavefunction in momentum space improves the results. 
\end{itemize}

\section{Variational forms mixing position and momentum spaces}
\begin{figure}[h]
    \centering
    \includegraphics[width=0.4\textwidth]{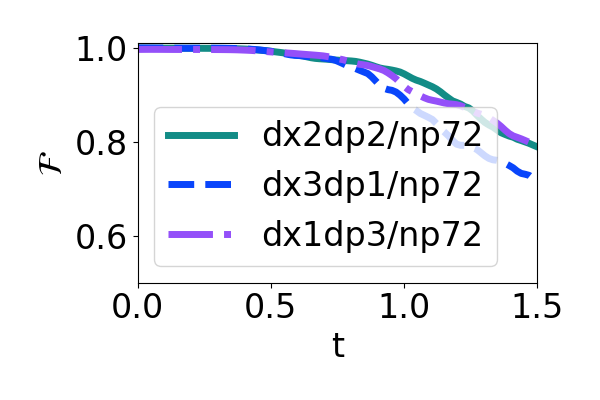}
    \caption{Fidelity of the variationally time-evolved wavefunction with respect to the exact one as a function of time. These results are obtained for the case of the Eckart barrier with 6 qubits. The quantum circuit corresponds to variational form vf1 with alternating layers in position and momentum space of depths given in the legend.}
    \label{fig:xpspace}
\end{figure}

In the main text, we discuss the improvement of the results when defining the \ansatz{} in the momentum space, \ie, by adding a \gls{qft} at the end of the circuit. 
Here, we show the results obtained by mixing momentum and position space in the variational form. 
More explicitly, the \ansatz{} is now composed of several parts; first, a part of given depth in position space, followed by an inverse \gls{qft}, then another piece of variational circuit with its own depth, and finally a \gls{qft} closing the circuit.
We refer to the part enclosed by the \glspl{qft} as the momentum part.
We vary the depths of the position and momentum parts. The results, shown in Fig.~\ref{fig:xpspace}, do not highlight any improvement in the performance. 

\section{Time evolution in presence of noise}

The robustness of variational approaches for quantum dynamics to hardware noise has been previously observed.~\cite{miessen2021} 
To evaluate the performance of our algorithm in presence of noise we modify the equations of motion as shown in Ref.~\cite{yuan2019} and work with the system's density matrix $\rho(t)$. The unitary evolution of mixed states defined by $\rho(t)$ under $\mathcal{H}$ is governed by the von Neumann equation 
\begin{equation}
    \frac{d\rho(t)}{dt} = -i [\mathcal{H}, \rho(t)].
\end{equation}
The equations of motion for the variational parameters then become
\begin{equation}
    \text{F}\dot{\bt}=\text{V} \ ,
\end{equation}
where
\begin{equation}
    F_{kj} := \text{Tr}\Bigg[ \bigg( \frac{\partial \rho(\bt)}{\partial \theta_k} \bigg)^{\dagger} \frac{\partial \rho(\bt)}{\partial \theta_j} \Bigg]
\end{equation}
and
\begin{equation}
    V_k := -i \text{Tr}\Bigg[ \bigg( \frac{\partial \rho(\bt)}{\partial \theta_k} \bigg)^{\dagger} [H,\rho(\bt)] \Bigg].
\end{equation}
In practice, $\rho(\bt)$ is extracted at each time step from a noisy quantum simulation performed with the \texttt{AerSimulator} (in \texttt{density\_matrix} mode) of \textsc{Qiskit}~\cite{Qiskit} with a given \texttt{NoiseModel}. 
The equations of motion are then solved classically with exact matrix/vector multiplications. 
This allows us to study the effect of particular types of hardware noise without adding the one of finite sampling on top. 
The fidelities, $\mathcal{F}(t)$, shown in this section are calculated as
\begin{equation}
    \mathcal{F}(t) = \Bigg( \text{Tr} \sqrt{\sqrt{\rho(t)}\rho_{\text{ex}}(t)\sqrt{\rho(t)}}\Bigg)^2
\end{equation}
where $\rho_{\text{ex}}(t) = \ket{\psi_{\text{ex}}(t)}\bra{\psi_{\text{ex}}(t)}$ with $\ket{\psi_{\text{ex}}(t)} = e^{-i\mathcal{H}t}\ket{\psi_0}$, the exact state at time $t$.

We select a system for which there is no numerical instabilities in the noiseless case, namely the free particle system discretized with 5 qubits with variational form vf2 depth 3 and initial width $\mathcal{B} = 3/\sqrt{2}$ (see Fig.~\ref{fig:sigmas}).
In all the following noisy simulations the state is initialized with the parameters obtained without noise, \ie, the same ones employed in the results shown in Section~\ref{sec:accuracy}, Fig.~\ref{fig:sigmas}. 

\begin{figure}
    \centering
    \includegraphics[width=0.4\textwidth]{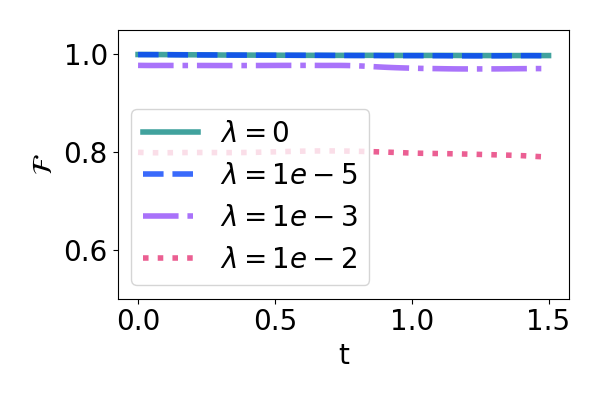}
    \caption{Fidelity of the variationally time-evolved density matrix with respect to the exact one as a function of time and in presence of depolarizing noise. These results are obtained for the case of the free particle with 5 qubits and variational form vf2.}
    \label{fig:fp_noise_dep}
\end{figure}

First, we apply a depolarizing channel implemented in \textsc{Qiskit}~\cite{Qiskit}, which is defined as
\begin{equation}
    E(\rho(t)) = (1-\lambda)\rho + \lambda \text{Tr}[\rho(t)]\frac{I}{2^{N_q}}
\end{equation}
where $N_q$ is the number of qubits and $\lambda$ is a parameter, $0 \leq \lambda \leq 4^{N_q} / (4^{N_q}-1)$.
We repeat the simulations for various values of $\lambda$ and show the results in Fig.~\ref{fig:fp_noise_dep}
Because the initial parameters are obtained from a noiseless optimization, the fidelity at $t=0$ does not lie above 99\% anymore when $\lambda$ increases. On the other hand, the fidelity does not significantly drop during the time evolution, suggesting that our algorithm is robust to a certain amount of noise. 

\begin{figure}
    \centering
    \includegraphics[width=0.4\textwidth]{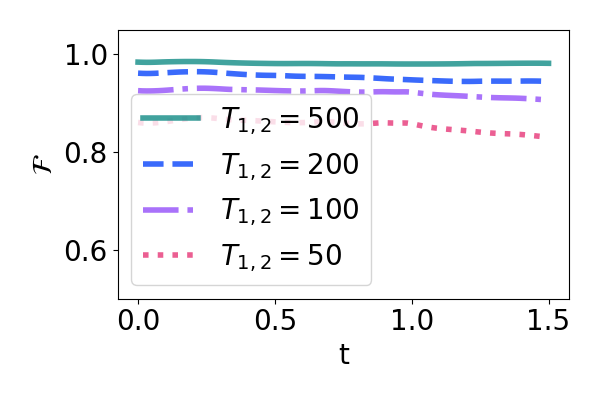}
    \caption{Fidelity of the variationally time-evolved density matrix with respect to the exact one as a function of time and in presence of thermal relaxation noise. The time constants $T_{1,2}$ are given in $\mu$s. These results are obtained for the case of the free particle with 5 qubits and variational form vf2.}
    \label{fig:tr_noise_dep}
\end{figure}

Second, we replace the depolarizing channel with thermal relaxation noise as implemented in \textsc{Qiskit}~\cite{Qiskit}. The operation time are set to 0 ns for the Rz gates, since they can be virtually applied by changing the phase of the subsequent pulses, and 36 ns and 500 ns for the Ry and CZ gates, respectively. 
The thermal relaxation time constant $T_1$ and the dephasing time constant $T_2$ are chosen to be equal and their value is varied from 50 $\mu$s to 500 $\mu$s (see Fig.~\ref{fig:tr_noise_dep}). 
In this case, we again only observe a small drop in accuracy during the dynamics for all noise regimes.
This strengthens the conjecture that variational time evolution approaches show good resilience to quantum hardware noise.

\clearpage
\section{Local diagonal space}

\begin{figure*}[h]
     \centering
     \begin{subfigure}[b]{0.4\textwidth}
         \centering
         \includegraphics[width=\textwidth]{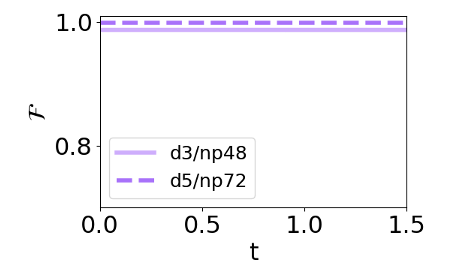}
         \caption{Harmonic oscillator}
         \label{fig:HO_ld}
     \end{subfigure}
     \begin{subfigure}[b]{0.4\textwidth}
         \centering
         \includegraphics[width=\textwidth]{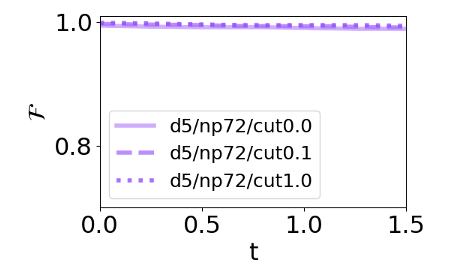}
         \caption{Eckart barrier}
         \label{fig:EB_ld}
     \end{subfigure}
        \caption{Fidelity, $\mathcal{F}$ as a function of time, $t$, of the \gls{vte} in LD space for (a) the harmonic oscillator, (b) the Eckart barrier. The results are obtained with 6 qubits and variational form vf2 at depth $d$ (with corresponding number $n_p$ of variational parameters). In the case of the Eckart barrier, the fidelity results are given for different cutoffs, $cut$, of the Hamiltonian coefficients used to obtain the diagonalization unitary. }
        \label{fig:ld_space}
\end{figure*}

In this section we discuss the addition of a unitary circuit for rotating the quantum state in a basis which partially diagonalizes the Hamiltonian. 
We first classically diagonalize the harmonic oscillator and Eckart barrier Hamiltonians. 
We then map the unitary matrix, made up of the eigenvectors sorted by increasing order of energy, to a quantum circuit using the isometry decomposition of Ref.~\cite{Iten2016} as implemented in \textsc{Qiskit}~\cite{Qiskit}. 
The resulting circuits, $\mathcal{D}$, are then appended to the variational form as
\begin{equation*}
    \Qcircuit @C=0.5em @R=0.5em @!R {
	 	\lstick{ \ket{0}^{\otimes N_q} } & \qw {/} & \qw & \gate{U(\bm{\theta})} & \gate{\mathcal{D}} & \meter  \\
	 	} \quad.
\end{equation*}
To obtain the results of Fig.~\ref{fig:ld_space}, we add $\mathcal{D}$ to the variational form vf2.
Those results show a high state fidelity over the entire simulation time. 
In the case of the Eckart potential, we also tested unitaries $\mathcal{D}'$, which only perform a partial diagonalization of the Hamiltonian.
More specifically, prior to diagonalization, we set to zero all matrix elements with absolute values below cutoff thresholds of $cut=0.1$ and $1.0$. We stress here that $cut$ should not be mixed up with $rc$ which is instead the reconditioning number used in the least-squares algorithm.
The density of non-zero elements in the resulting Hamiltonian matrix is then $0.98$, $0.27$, and $0.14$ for cutoffs $0$, $0.1$, and $1.0$, respectively.
The results exhibit high accuracy for all cutoff values as seen from Fig.~\ref{fig:EB_ld}.

In the following we give insight on the reduction of the number of necessary variational parameters when working in the local diagonal space. 
The aim of the local diagonalization is to find a unitary transformation of the form $\mathcal{D} = \bigotimes_{i=1}^M\mathcal{D}_i$ which captures a great part of the dynamics, and turns the variational problem into a smooth dynamical problem.
This can be achieved by choosing $\mathcal{D}$ such as 
\begin{equation}
    \mathcal{D} : \ket{x_n} \xrightarrow{} \ket{\Phi_n^0}
\end{equation}
where the $\ket{\Phi_n^0}$ are the eigenstates, associated to eigenvalues $E_n^0$, of the simpler Hamiltonian $\mathcal{H}_0$ related to the original problem's Hamiltonian via a perturbation $\tilde{\mathcal{V}}$ as
\begin{equation}
    \mathcal{H} = \mathcal{H}_0 + \tilde{\mathcal{V}}.
\end{equation}
Note that $\mathcal{H}_0$ can be adapted along the dynamics \eg, could be defined as a time-dependent mean-field Hamiltonian.\\
For the sake of clarity, let us first consider a variational \ansatz{} defined in the basis of the eigenstates $\ket{\Phi_n^0}$ and with the variational parameters, $\alpha_n$, simply being the amplitudes associated with those eigenvectors:
\begin{equation}
    \ket{\psi(\bm{\alpha})} = \sum_{n=1}^{N} \alpha_n e^{-iE_n^0t} \ket{\Phi_n^0}.
    \label{eq:easy_ansatz}
\end{equation}
Clearly, under the above assumption the parameters are expected to vary smoothly in time with the limit $\dot{\alpha}_n \xrightarrow{} 0 $ for $\tilde{\mathcal{V}} \xrightarrow{} 0$. 
Moreover, since $\mathcal{H}_0$ is expected to describe $\mathcal{H}$ reasonably well, the energies $E_n^0$ are close to the $E_n$ (eigenvalues of $\mathcal{H}$). Because of the conservation of energy, the dynamics is approximately confined to an invariant subspace $\mathbb{A}$ (defined by the initial state). In other words, the $\ket{\Phi_n^0}$ lying far in energy from the eigenstates spanning $\mathbb{A}$ are very unlikely to enter the variational wavefunction at any point during the dynamics. 
The subspace $\mathbb{A}$ is only spanned by a limited number of eigenvectors of $\mathcal{H}_0$, $\mathbb{A} = \text{span}\{ \ket{\Phi_n}\}_{n\in I_{\mathbb{A}}} $, for a subset $I_{\mathbb{A}}$ of indices $n$. 
Hence, a considerable reduction of complexity is apparent, since one can decrease the number of parameters by neglecting the $\alpha_n$ for which $n \notin I_{\mathbb{A}}$. To stay on the safe side, one can use a sector $\mathbb{A}$ of the Hilbert space that is larger than the one strictly needed to describe the initial state, and that (in the language of perturbation theory) can accommodate the most important virtual transitions. 
Even in this case, the bottom line remains unaltered: a reduction of complexity is possible since $\mathcal{H}_0$ captures the essence of the dynamics, hence its eigenspaces are \textit{quasi}-invariant under the action of $\mathcal{H}$. 
Although our problem is not in the form of Eq.~\ref{eq:easy_ansatz}, since our $\theta_k$ are not directly the amplitude associated to a given $\ket{\Phi_k^0}$ but real parameters entering the variational \ansatz{}, $\ket{\psi(\bt)} = \mathcal{U}(\bt)\ket{\phi}$, the same observation applies.
Our variational parameters only need to describe subspace $\mathbb{A}$.
As a result, the number of parameters, $n_p$, must be on the order $2\,\text{dim}(\mathbb{A})$ to fulfill the requirement that for each $\ket{\Phi} \in \mathbb{A}$ there exists a set of parameter $\bt$  such that
\begin{equation}
    \mathcal{F}(\bt) = |\braket{\Phi|\psi(\bt)}|^2 \approx 1.
\end{equation}
There can be some flexibility in $n_p$ depending on the target accuracy for the state representation.
In any case, $n_p$ now scales with $\text{dim}(\mathbb{A})$ leading to a reduction of the overall complexity.

\end{document}